\documentclass[sigconf]{acmart}

\newcommand{\phib}{\boldsymbol \phi}

\newcommand{\Pb}{{\boldsymbol P}}

\newcommand{\Xb}{{\boldsymbol X}}

\newcommand{\bb}{{\boldsymbol b}}

\newcommand{\grad}{{\nabla}}

\newcommand{\deformMap}{\phib}

\newcommand{\solidDen}{R}
\newcommand{\pkstress}{\Pb}
\newcommand{\bodyforce}{\bb}

\newcommand{\bX}[0]{\boldsymbol{X}}
\newcommand{\bx}[0]{\boldsymbol{x}}
\newcommand{\xhatn}[0]{\hat{\boldsymbol{x}}_n}
\newcommand{\xhat}[0]{\hat{\boldsymbol{x}}}

\newcommand{\eeg}[1]{}
\newcommand{\mml}[1]{}
\usepackage{tablefootnote}
 \usepackage{amsmath}

\DeclareMathOperator*{\argmin}{arg\,min}
\usepackage{cleveref}

\AtBeginDocument{%
  \providecommand\BibTeX{{%
    \normalfont B\kern-0.5em{\scshape i\kern-0.25em b}\kern-0.8em\TeX}}}

\acmSubmissionID{637}

\copyrightyear{2023}
\acmYear{2023}
\setcopyright{rightsretained}
\acmConference[SA Conference Papers '23]{SIGGRAPH Asia 2023 Conference Papers}{December 12--15, 2023}{Sydney, NSW, Australia}
\acmBooktitle{SIGGRAPH Asia 2023 Conference Papers (SA Conference Papers '23), December 12--15, 2023, Sydney, NSW, Australia}
\acmDOI{10.1145/3610548.3618207}
\acmISBN{979-8-4007-0315-7/23/12}

\begin{document}

\title{Neural Stress Fields \\for Reduced-order Elastoplasticity and Fracture}
\author{Zeshun Zong}
\affiliation{
\institution{UCLA}
\country{USA}
}
\orcid{0000-0002-3256-1692}

\author{Xuan Li}
\affiliation{
\institution{UCLA}
\country{USA}
}
\orcid{0000-0003-0677-8369}

\author{Minchen Li}
\affiliation{
\institution{UCLA}
\institution{Carnegie Mellon University}
\country{USA}
}
\orcid{0000-0001-9868-7311}

\author{Maurizio M. Chiaramonte}
\affiliation{
\institution{Meta Reality Labs Research}
\country{USA}
}
\orcid{0000-0002-2529-3159}

\author{Wojciech Matusik}
\affiliation{
\institution{MIT CSAIL}
\country{USA}
}
\orcid{0000-0003-0212-5643}

\author{Eitan Grinspun}
\affiliation{
\institution{University of Toronto}
\country{Canada}
}
\orcid{0000-0003-4460-7747}

\author{Kevin Carlberg}
\affiliation{
\institution{Meta Reality Labs Research}
\country{USA}
}
\orcid{0000-0001-8313-7720}

\author{Chenfanfu Jiang}
\affiliation{
\institution{UCLA}
\country{USA}
}
\orcid{0000-0003-3506-0583}

\author{Peter Yichen Chen}
\affiliation{
\institution{MIT CSAIL}
\country{USA}
}
\orcid{0000-0003-1919-5437}

\renewcommand{\shortauthors}{Z. Zong, X. Li, M. Li, M. Chiaramonte, W. Matusik, E. Grinspun, K. Carlberg, C. Jiang, P. Chen.}

\begin{abstract}
We propose a hybrid neural network and physics framework for reduced-order modeling of elastoplasticity and fracture. State-of-the-art scientific computing models like the Material Point Method (MPM) faithfully simulate large-deformation elastoplasticity and fracture mechanics. However, their long runtime and large memory consumption render them unsuitable for applications constrained by computation time and memory usage, e.g., virtual reality. To overcome these barriers, we propose a reduced-order framework. Our key innovation is training a low-dimensional manifold for the Kirchhoff stress field via an implicit neural representation. This low-dimensional neural stress field (NSF) enables efficient evaluations of stress values and, correspondingly, internal forces at arbitrary spatial locations. In addition, we also train neural deformation and affine fields to build low-dimensional manifolds for the deformation and affine momentum fields. These neural stress, deformation, and affine fields share the same low-dimensional latent space, which uniquely embeds the high-dimensional simulation state. After training, we run new simulations by evolving in this single latent space, which drastically reduces the computation time and memory consumption. Our general continuum-mechanics-based reduced-order framework is applicable to any phenomena governed by the elastodynamics equation. To showcase the versatility of our framework, we simulate a wide range of material behaviors, including elastica, sand, metal, non-Newtonian fluids, fracture, contact, and collision. We demonstrate dimension reduction by up to 100,000$\times$ and time savings by up to 10$\times$.
\end{abstract}



\begin{teaserfigure}
  \includegraphics[width=\textwidth]{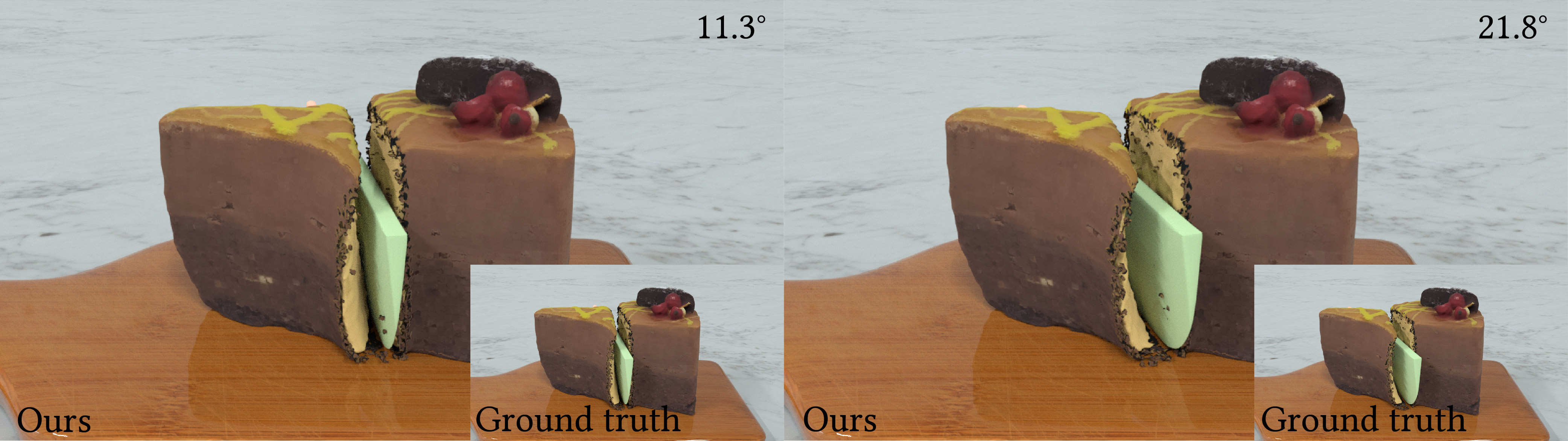}
  \caption{Our reduced-order model accurately simulates the cutting of a chocolate cake at various angles by time-stepping in a latent space of only dimension $r=6$. The original full-order simulation employs $200,000$ particles. Courtesy of dimension reduction, our approach is 10.2$\times$ faster than the full-order simulation.}
  \label{fig:cake_placeholder}
\end{teaserfigure}

\maketitle

\section{Introduction}
Physical simulation plays a crucial role in computational mechanics, digital twins, computational design, robotics, animation, visual effects, and virtual reality. A crucial class of these physical simulations are those governed by the conservation of momentum equation \citep{gonzalez2008first},
\begin{align}
    \solidDen\ddot{\deformMap} = \grad^{\Xb} \cdot \pkstress + \solidDen\bodyforce, \label{pde}
\end{align}
where $\pkstress$ is the first Piola-Kirchhoff stress, $\deformMap$ is the deformation map, $\solidDen$ is the initial density, $\bodyforce$ is the body force, and $\boldsymbol{X}\in \Omega_0$ is the reference position defined over domain $\Omega_0$. This partial differential equation (PDE) governs a wide range of elastoplastic behaviors.

To numerically solve this PDE, one has to spatially and temporally discretize it, e.g., via finite difference, finite element, or finite volume methods. A particularly flexible discretization framework is the material point method (MPM) \citep{sulsky1995application,jiang2016material}. MPM discretizes the spatial field via both Lagrangian particles and Eulerian grids. Thanks to this dual discretization paradigm, MPM thrives at handling large deformations, topology changes, and self-contact. 

Nevertheless, MPM's versatility also comes at the cost of computation burden, in terms of both long runtime and excessive memory consumption. To obtain accurate results, MPM tracks a large number of state variables through the particles, often at the order of millions. Such a computation bottleneck significantly hinders the feasibility of deploying MPM in time-critical and memory-bound applications. Notably, MPM's high-dimension state variables also pose a challenge in applications where synchronization is required. For example, in virtual reality and cloud gaming, multiple users share the same simulated physical environment; each user's simulation state needs to be efficiently shared with others via internet streaming. Synchronizing millions of MPM particle data at frame rate is simply not possible.

We propose to solve these computational challenges via reduced-order modeling (ROM), also known as model reduction \citep{barbivc2005real}. ROM reduces the computation cost by training a low-dimensional latent embedding of the original high-dimensional simulation data. After training, instead of evolving the original high-dimensional state variables over time, ROM only needs to time-step in the low-dimensional latent space, and synchronization between users only requires sharing the low-dimensional latent vector. The classic reduced-order, elasticity-only finite element method (FEM) \citep{sifakis2012fem} trains a low-dimensional embedding for the (discretized) deformation map $\deformMap$ in \cref{pde}. However, the low-dimensional deformation embedding alone is not enough for MPM and elastoplasticity simulations in general. 

\paragraph{History-dependent plasticity state variables.} MPM simulations feature history-dependent effects, e.g., plastic deformations of sands or metals. The low-dimensional deformation embedding by itself is unable to determine the plasticity state variables that are crucial for MPM time-stepping.

\paragraph{Deformation gradients as independent state variables.} MPM treats the deformation \emph{gradient} as a separate state variable that evolves independently from deformation state variables. Again, the low-dimensional deformation embedding cannot capture these deformation gradients.

Our key observation is that the ultimate purpose of \emph{all} these additional state variables is computing the stress field $\pkstress$ in \cref{pde}. As such, we can bypass the need to capture these intermediate state variables by directly training a low-dimensional embedding for the stress field itself. The low-dimensional stress and deformation embedding together capture all the information necessary for MPM time-stepping. We construct the low-dimensional stress embeddings via implicit neural representations, also known as neural fields. Our neural stress field (NSF)
approach enables stress evaluation and, in turn, force evaluation at arbitrary spatial locations. In a similar vein, we build low-dimensional neural deformation fields. To support MPM's affine particle-in-cell transferring scheme \citep{jiang2015affine}, we also build low-dimensional neural affine fields for the affine momentum field. All these three neural fields share the same latent space.

After training, we solve new physical simulation problems via projection-based latent space dynamics \citep{benner2015survey,carlberg2017galerkin}. During this PDE-constrained latent space dynamics stage, we obtain computation savings by evaluating the neural fields only at a small spatial subset, similar to the idea of cubature \cite{an2008optimizing}.
Our general, stress-based ROM approach works with any problem governed by the momentum equation \cref{pde}. To showcase the versatility of our approach, we validate NSF on a wide range of elastoplastic phenomena, including elastica, fracture, metal, sand, non-Newtonian fluids, contact, and collision. We demonstrate dimension reduction of 100,000$\times$ and computation savings of 10$\times$.

\section{Related Work}
\label{sec:related work}    
\paragraph{The Material Point Method}
\citet{sulsky1995application} introduced MPM by combining Lagrangian and Eulerian techniques for solid mechanics, drawing upon the earlier works by \citet{harlow1962particle,brackbill1986flip} on PIC/FLIP. Since its introduction to the graphics community \citep{hegemann2013level,stomakhin2013material}, MPM has garnered considerable attention. Its primary advantage in modeling elastoplastic materials lies in its capability to handle extreme deformation and topological changes. MPM has been successfully applied to simulate various phenomena, including granular media \citep{klar2016drucker,daviet2016semi,yue2018hybrid,chen2021hybrid}, non-Newtonian fluids \citep{yue2015continuum,fei2019multi}, viscoelasticity \citep{fang2019silly}, fracture \citep{wolper2019cd,wang2019simulation,wolper2020anisompm}, and thermomechanics \citep{ding2019thermomechanical}. Efforts have been made to speed up MPM simulations through GPU \citep{gao2018gpu,wang2020massively,fei2021principles}, multi-node \citep{qiu2023sparse}, and multigrid \citep{wang2020hierarchical} accelerations, as well as compiler optimization \citep{hu2019taichi}. However, the substantial computational cost and memory consumption of MPM still present challenges that need to be addressed.
\paragraph{Reduced-order Modeling}
Classic reduced-order modeling methods employ linear subspaces \citep{barbivc2005real,sifakis2012fem}. These subspaces are often constructed via principal component analysis and, equivalently, proper orthogonal decomposition \citep{berkooz1993proper,holmes2012turbulence}. These linear subspaces have been successively applied to solids \citep{an2008optimizing,kim2009skipping,barbivc2011real,yang2015expediting,xu2015interactive} and fluids \citep{treuille2006model,kim2019deep,kim2013subspace,wiewel2019latent}. Recently ROM methods have been exploring nonlinear low-dimensional manifolds, often leveraging autoencoder neural networks \citep{lee2020model}. These nonlinear approaches enable smaller latent space dimensions in comparison with the classic linear approaches \citep{fulton2019latent,shen2021high}. Our technique also falls into this nonlinear model reduction category.

Relatedly, there has been lots of progress in data-driven latent space dynamics \citep{lusch2018deep}, and the entire latent space evolution is strictly learned via another neural network, e.g., recurrent neural networks \citep{wiewel2019latent}. By contrast, our method follows the classic, invasive ROM literature and evolves the latent space using the numerical methods and PDEs that were used to generate the training data. In our method, the latent space dynamics are entirely PDE-based without any data-driven component.
\paragraph{Neural Fields} A neural field \citep{xie2021neural} parameterizes a spatially dependent vector field via a neural network. The pioneering works by \citet{park2019deepsdf, chen2019learning,mescheder2019occupancy} employ this representation for signed distance fields, where different latent space vector corresponds to different geometries. Since then, it has been widely adopted for neural rendering \citep{mildenhall2020nerf}, topology optimization \citep{zehnder2021ntopo}, geometry processing \citep{yang2021geometry,aigerman2022neural}, and various PDE problems \citep{raissi2019physics,chen2022implicit}. Recently, \citet{pan2022neural,chen2023model,chen2023crom} have leveraged neural fields for ROM. Notably, \citet{chen2023model} build a neural-field-based, reduced-order framework for MPM. Their approach constructs a low-dimensional embedding only for the deformation field. Consequently, their method is unable to handle history-dependent plasticity, and the deformation gradient computed from differentiating the learned deformation field is too inaccurate for large deformation phenomena such as a fracture. As a major point of departure, we train a low-dimensional manifold directly for the stress field and can therefore handle both plasticity and fracture. Furthermore, we achieve angular momentum conservation by training a low-dimensional neural affine field while \citep{chen2023model}'s formulation suffers from excessive dissipation. 

\section{Background: full-order MPM}
This section will briefly recap the essential ingredients of the full-order MPM model. \Cref{sec:reduce:kinematics,sec:reduce:dynamics} will introduce the corresponding reduced-order model. We refer to \citet{sulsky1995application,jiang2016material} for additional MPM details.

\subsection{Finite strain elasticity and elastoplasticity}
Let $\Omega_0 \subset \mathbb{R}^3$ denote the material space and $\Omega_t$ the world space at time $t.$ We are interested in the dynamics of a continuum in time $t \in [0, T].$ The deformation map $\boldsymbol{x} := \boldsymbol{\phi}(\boldsymbol{X},t)$ maps $\boldsymbol{X}\in \Omega_0$ to world space coordinate $\boldsymbol{x} \in \Omega_t.$ From the Lagrangian view, the dynamics of a continuum can be described by a density field $R(\boldsymbol{X}, t): \Omega_0 \times [0,T] \rightarrow \mathbb{R}$ and a velocity field $\boldsymbol{V}(\boldsymbol{X}, t)=\frac{\partial \boldsymbol{\phi}(\boldsymbol{X}, t)}{\partial t}: \Omega_0 \times [0, T] \rightarrow \mathbb{R}^3.$ They are governed by the conservation of mass 
\begin{equation}
    R(\boldsymbol{X}, t) J(\boldsymbol{X}, t)=R(\boldsymbol{X}, 0),
    \label{eqn:conserve_mass}
\end{equation}
 and the conservation of momentum
 \begin{equation}
     R(\boldsymbol{X}, 0) \frac{\partial \boldsymbol{V}}{\partial t}(\boldsymbol{X}, t)=\nabla^{\boldsymbol{X}} \cdot \boldsymbol{P}+R(\boldsymbol{X}, 0) \boldsymbol{g}.
     \label{eqn: conserve_momentum}
 \end{equation}

Here $J=\operatorname{det} (\boldsymbol{F}),$  $\boldsymbol{F}=\frac{\partial \boldsymbol{\phi}}{\partial \boldsymbol{X}}(\boldsymbol{X}, t)$ is the deformation gradient, $\boldsymbol{P}$ is the first Piola-Kirchhoff stress, and $\boldsymbol{g}$ is the gravity term. $\boldsymbol{P}$ can be related to the Kirchhoff stress $\boldsymbol{\tau}$ as $\boldsymbol{P} = \boldsymbol{\tau} \boldsymbol{F}^{-T}.$ 

For a hyperelastic solid, the Kirchhoff stress can be computed as $\boldsymbol{\tau} = \frac{\partial \psi}{\partial \boldsymbol{F}}{(\boldsymbol{F})} \boldsymbol{F}^T,$ where $\psi$ is the energy density function of the chosen constitutive model. For an elastoplastic continuum, the deformation gradient is multiplicatively decomposed into $\boldsymbol{F}=\boldsymbol{F}^E \boldsymbol{F}^P,$ with the former being the elastic deformation that supplies elastic force, and the latter being the permanent plastic deformation gradient. The decomposition requires that $\boldsymbol{\tau}(\boldsymbol{F}^E)$ lies within an admissible region defined by some yield condition $y(\boldsymbol{\tau})<0.$ Given $\boldsymbol{F},$ $\boldsymbol{F}^E$ evolves from $\boldsymbol{F},$ following some plastic flow until the yield condition is satisfied. The procedure is often called return mapping. 

\subsection{MPM discretization}
MPM discretizes a continuum bulk into a set of Lagrangian particles $p,$ and discretizes time $t$ into a sequence of timesteps $t_0 = 0, t_1, t_2, ...$ Here we take a fixed stepsize $\Delta t,$ so $t_n = n \Delta t.$ The advection is performed on particles so \cref{eqn:conserve_mass} is naturally satisfied. If we approximate $\boldsymbol{V}^n$ by $\frac{1}{\Delta t} (\boldsymbol{X}^{n+1} - \boldsymbol{X}^n),$ and assume no gravity and free surface for clarity, for an arbitrary test function $Q,$ the weak form of \cref{eqn: conserve_momentum} is then given by 
\begin{equation}
    \label{eqn:weak_form}
    \int_{\Omega_0} R(\boldsymbol{X}, 0)\frac{1}{\Delta t} \left(\boldsymbol{V}^{n+1}-\boldsymbol{V}^n\right) Q d \boldsymbol{X}=-\int_{\Omega_0} \boldsymbol{P} \nabla^{\boldsymbol{X}} Q d \boldsymbol{X}.
\end{equation}

Pushing forward the integral from $\Omega_0$ to $\Omega_{n} = \Omega_{t_n},$ we obtain 
\begin{equation}
\label{eqn:weak_form_omegat}
    \int_{\Omega_n} \rho\left(\boldsymbol{x}, t^n\right)\frac{1}{\Delta t} \left({\boldsymbol{v}}^{n+1}-\boldsymbol{v}^n\right) q d \boldsymbol{x}=-\int_{\Omega_n} \frac{1}{J^n} \boldsymbol{P} \boldsymbol{F}^{n T} \nabla^{\boldsymbol{x}} q d \boldsymbol{x},
\end{equation}
where $\rho, \boldsymbol{v}^n, \boldsymbol{v}^{n+1}$ and $q$ are the Eulerian counterparts of $R, \boldsymbol{V}^n, \boldsymbol{V}^{n+1}$ and $Q,$ respectively \citep{jiang2016material}.

MPM adopts B-Spline-based interpolations and uses material particles $p$ as quadratures to approximate the integration \cref{eqn:weak_form_omegat}. Let $m_p$ denote the mass of particle $p$ with initial position $\boldsymbol{X}_p.$ Denote its position and velocity at time $t_n$ by $\boldsymbol{x}_p^n$ and $\boldsymbol{v}_p^n.$ Let $m_i$ and $\boldsymbol{v}_i$ denote the mass and velocity on background grid node $i$ at position $\boldsymbol{x}_i.$ Let $N(\boldsymbol{x})$ denote the weight function, and $w_{ip}^n = N\left(\boldsymbol{x}_p^n-\boldsymbol{x}_i\right).$ Employing mass lumping, we can express the force equilibrium as 
\begin{equation}
    \label{eqn:node force balance}
    \frac{1}{\Delta t} m_i^n\left({\boldsymbol{v}}_i^{n+1}-\boldsymbol{v}_i^n\right)=-\sum_p \boldsymbol{\tau}_p^n \nabla w_{i p}^n V_p^0,
\end{equation}
thus providing a way to update the next stage grid velocities $\boldsymbol{v}_i^{n+1}.$ Here $V_p^0$ and $\boldsymbol{\tau}_p^n$ are the initial volume and Kirchhoff stress at time $t_n$ of material particle $p.$
\subsection{MPM algorithm}
\label{sec:mpm algo}
At each step, particle mass and momentum are transferred to grid nodes. Grid velocities are updated and then transferred back to particles for advection. Let $\boldsymbol{C}_p^n$ denote the affine momentum of particle $p$ at time $t_n.$ The explicit MPM algorithm can therefore be summarized as the following: 

\begin{enumerate}
    \item \textbf{P2G.} Transfer mass and momentum from particles to grid as $m_{{i}}^n=\sum_p w_{{i} p}^n m_p$ and $m_i^n\boldsymbol{v}_{{i}}^n= \sum_p w_{{i} p}^n m_p\left(\boldsymbol{v}_p^n+\boldsymbol{C}_p^n\left(\boldsymbol{x}_{{i}}-\boldsymbol{x}_p^n\right)\right),$ if the APIC transfer scheme is adopted. If the conventional PIC scheme is adopted, the latter is simply replaced by $m_i^n\boldsymbol{v}_{{i}}^n= \sum_p w_{{i} p}^n m_p\boldsymbol{v}_p^n.$
    \item \textbf{Grid update.} Update grid velocities at next timestep by $\boldsymbol{v}_i^{n+1} = \boldsymbol{v}_i^{n} - \frac{\Delta t}{m_i} \sum_p \boldsymbol{\tau}_p^n \nabla w_{i p}^n V_p^0 + \Delta t \boldsymbol{g}.$ Collision and Dirichlet boundary conditions are also handled at this stage.
    \item \textbf{G2P.} Transfer velocities back to particles and update particle states. $\boldsymbol{v}_p^{n+1}=\sum \boldsymbol{v}_{{i}}^{n+1} w_{{i} p}^n,$ $\boldsymbol{x}_p^{n+1} = \boldsymbol{x}_p^{n} + \Delta t \boldsymbol{v}_p^{n+1},$ $${\boldsymbol{C}}_p^{n+1}=\frac{12}{\Delta x^2(b+1)} \sum_{{i}} w_{{i} p}^n \boldsymbol{v}_i^{n+1} \left(\boldsymbol{x}_{{i}}^n-\boldsymbol{x}_p^n\right)^T,$$
    $$\boldsymbol{F}_p^{\text{trial}, n+1} =\left(\mathbf{I}+\Delta t \sum_i \boldsymbol{v}_i^{n+1}\left(\nabla w_{i p}^n\right)^T\right) \boldsymbol{F}_p^{E,n},$$
    $$\boldsymbol{F}_p^{E,n+1} = \text{returnMap}(\boldsymbol{F}_p^{\text{trial}, n+1}) \text{ and } \boldsymbol{\tau}_p^{n+1} = \boldsymbol{\tau}(\boldsymbol{F}_p^{E,n+1}).$$ Here $b$ is the B-spline degree, and $\Delta x$ is the Eulerian grid spacing. If additional damping is desired, RPIC can be added in the computation of ${\boldsymbol{C}}_p$ as in \citep{fang2019silly}.
\end{enumerate}
\section{Reduced-order model: kinematics}
\label{sec:reduce:kinematics}
\begin{figure}[h]
  \centering
  \includegraphics[width=\linewidth]{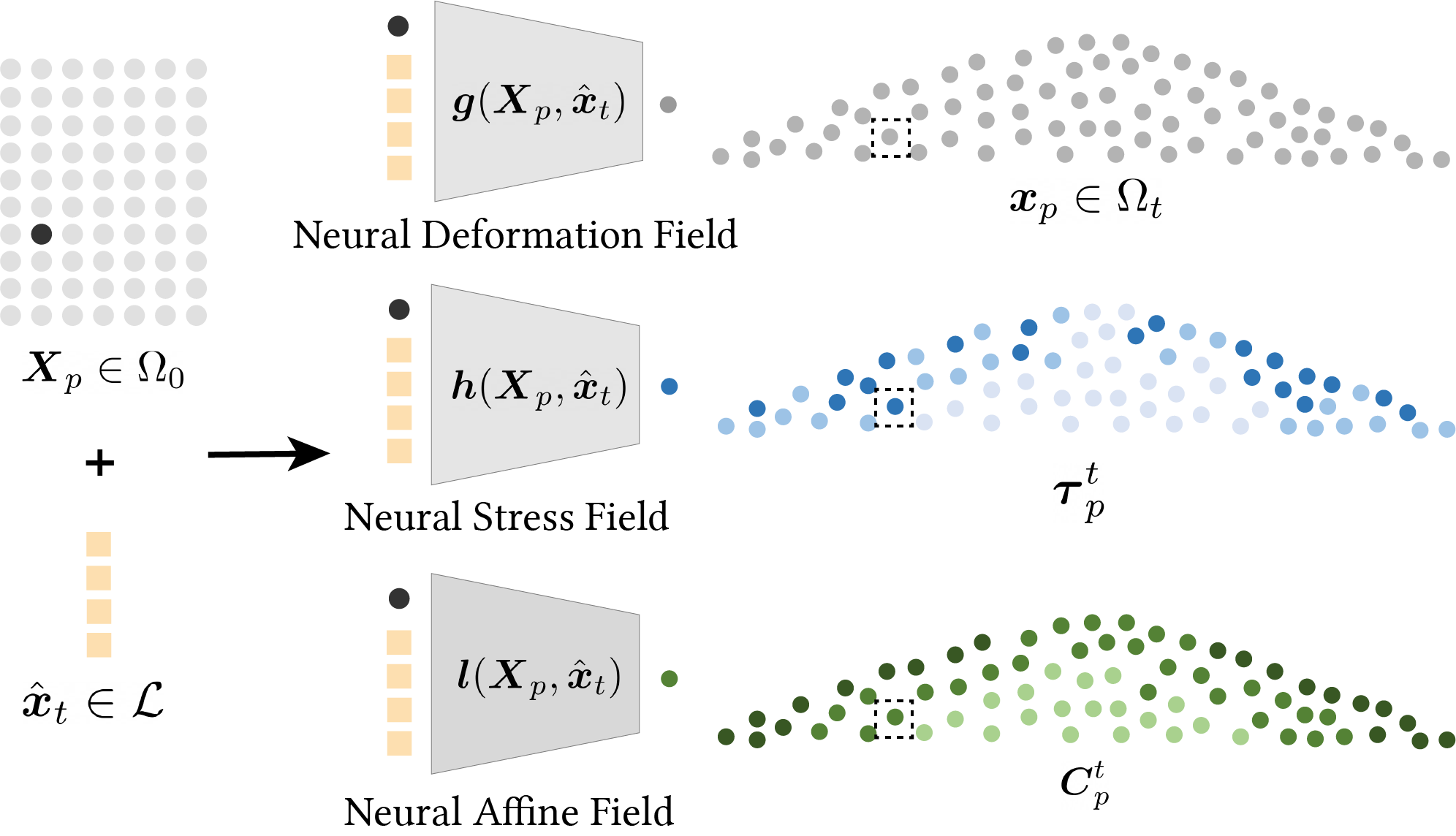}
  \caption{Latent space kinematics. Given a latent space vector $\xhat_t \in \mathcal{L},$ evaluating the neural deformation, stress, and affine fields at \emph{any} reference position $\bX_p \in \Omega_0$ (e.g., the black dot in $\Omega_0$) results in the corresponding deformation, stress, and affine momentum at time $t$ at the current position (the boxed dot in $\Omega_t$).}
  \Description{}
  \label{fig:latent_kinematics}
\end{figure}

To reduce the full-order MPM model, we will construct a nonlinear approximation to the solution of \cref{eqn: conserve_momentum} over a low-dimensional manifold. A schematic illustration is shown in \cref{fig:latent_kinematics}.

\subsection{Low-dimensional Manifold Construction}
Let the continuous field $\boldsymbol{f}(\boldsymbol{X},t; \mu): \Omega_0\times [0,T] \rightarrow \mathbb{R}^m$ denote any relevant state variable in the solution to \cref{eqn: conserve_momentum} for $\boldsymbol{X}\in \Omega_0$ at time $t$. Example state variables include the deformation map, stress, etc. 
Here, $\mu$ is the generalized problem parameter, including but not limited to material parameters, initial conditions, and boundary conditions. Choice of $\mu$ for each experiment will be detailed in \cref{sec:experiments}. We seek a continuous field ${\boldsymbol{\hat{f}}}(\cdot; \xhat)$ defined over $\Omega_0$ and parameterized by $\xhat \in \mathcal{L},$ a low-dimensional latent space, such that 
\begin{equation}
    {\boldsymbol{\hat{f}}}(\bX; \xhat(t, \mu)) \approx \boldsymbol{f}(\boldsymbol{X},t; \mu),  \forall \boldsymbol{X} \in \Omega_0 \text{ and } \forall t \in [0,T].
\end{equation}
The dimension $r$ of $\mathcal{L}\subset \mathbb{R}^r$ is taken to be a small number so that the dynamics of a continuum becomes the evolution of the latent space vector $\xhat$ in a low-dimensional latent space $\mathcal{L}.$ For notational simplicity, we will omit explicit dependence on $\mu$. To computationally construct any of these low-dimensional manifolds, we will employ a neural field, also known as implicit neural representation \citep{xie2021neural}. Next, we will discuss specific MPM state variables for which we will build neural fields.

\subsection{Neural Deformation Fields}
Similar to classic elastic-only FEM, one must build a low-dimensional manifold for the deformation field $\boldsymbol{\phi}(\bX, t)$ \citep{barbivc2005real}. We achieve this by constructing a manifold $\boldsymbol{g}(\bX, \xhat)$ \citep{chen2023model} such that
\begin{equation}
    \forall \bX \in \Omega_0, \forall t \in [0,T], \boldsymbol{g}(\bX, \xhat_t) \approx \boldsymbol{\phi}(\bX, t) = \bx_t.
    \label{eqn:approx_def_map}
\end{equation}

\subsection{Neural Stress Fields}
Unlike elasticity-only FEM, MPM features additional history-based plastic effects and state variables. Moreover, the deformation gradient $\boldsymbol{F}$ is treated as an evolving state variable independent of $\bx$. To address these various state variables, we observe that representing the stress field is a neat yet effective approach. Since the eventual goal of \emph{all} these state variables is computing the stress tensor, by directly building a low-dimensional manifold for the stress tensor, we avoid cumbersome treatment of numerous plasticity state variables as well as inaccurate calculation of deformation gradient. We approximate the Kirchhoff stress field $\boldsymbol{\tau}(\bx,t)$ by a manifold $\boldsymbol{h}(\bX, \xhat)$ such that 
\begin{equation}
    \forall \bX \in \Omega_0, \forall t \in [0,T], \boldsymbol{h}(\bX, \xhat_t) \approx \boldsymbol{\tau}(\bx,t) = \boldsymbol{\tau}(\boldsymbol{\phi}(\bX,t), t).
\end{equation}
The right hand side of \cref{eqn:weak_form_omegat} can thus be approximated as $$-\int_{\Omega_n}\frac{1}{J^n} \boldsymbol{\tau}(\boldsymbol{x}) \nabla^{\boldsymbol{x}} q d\bx = -\int_{\Omega_0} \boldsymbol{\tau}(\bX) \nabla^{\boldsymbol{x}} q d\bX \approx -\int_{\Omega_0} \boldsymbol{h}(\bX, \xhat_n)\nabla^{\boldsymbol{x}} q d\bX,$$ which naturally fits within the spatial discretization of MPM.

\paragraph{Remarks}
(1) An alternative approach is to use the deformation gradient to compute the stress. The deformation gradient can be computed by differentiating the neural deformation field \citep{chen2023model}. However, the numerical deformation gradient $\boldsymbol{F}_{\text{MPM}}$ in the full-order MPM is not computed from $\frac{\partial\boldsymbol{\phi}}{\partial \bX},$ but rather numerically integrated. Consequently, this approach will cease to provide accurate grid forces when $\frac{\partial\boldsymbol{\phi}}{\partial \bX}$ does not resemble $\boldsymbol{F}_{\text{MPM}}$, e.g., in numerical fracture. A well-trained neural stress field, on the other hand, directly supplies the correct grid forces for MPM grid update. (2) Since stress is computed from the elastic part of the deformation gradient $\boldsymbol{F}^{E} = \text{returnMap}(\boldsymbol{F}^{\text{trial}}),$ the plastic flow is implicitly stored. Evaluation of the return map can be avoided in deployment, thus reducing the computational cost. Overall, our neural-stress-field approach is a general approach that allows for reduced-order solutions for all the standard plasticity models.

\subsection{Neural Affine Fields}
Additionally, to accommodate for the affine momentum term $\boldsymbol{C}$ used in APIC and RPIC transfer scheme (\cref{sec:mpm algo}), we construct another manifold $\boldsymbol{l}(\bX, \xhat)$ such that $\boldsymbol{l}(\bX, \xhat_n) \approx \boldsymbol{C}(\boldsymbol{\phi}(\bX,t),t).$ This field enables angular momemtum conservation \citep{jiang2015affine}. \eeg{using 
$\hat{\boldsymbol{C}}(\bX, \xhat_n) \approx \boldsymbol{C}(\boldsymbol{\phi}(\bX,t),t)$ would avoid memorizing a relationship? }

\subsection{Network training}
Let $\mathcal{T} = \{t_0, t_1, ..., t_N = T\},$ $\mathcal{P}$ denote the set of all material particles $p,$ $\mathcal{U}$ denote the set of problem parameters $\mu$ that we are interested in, and $\mathcal{U}_{\text{train}} \subset \mathcal{U}$ a subset for training. Let the training set be $\{(\bx_p^n, \boldsymbol{\tau}_p^n, \boldsymbol{C}_p^n): p\in \mathcal{P}, \mu \in \mathcal{U}_{\text{train}} \}$. Define $\boldsymbol{x}^n = [\boldsymbol{x}_1^n, \boldsymbol{x}_2^n, ..., \boldsymbol{x}_{|\mathcal{P}|}^n]^T.$
The implementation of the three manifolds is summarized below:
\begin{enumerate}
    \item Train displacement decoder network $\boldsymbol{g}_{\theta_g}(\bX, \hat{\boldsymbol{x}})$ and encoder network $\boldsymbol{e}_{\theta_e}(\boldsymbol{x}^n)$ by 
    $$\min_{\theta_g, \theta_e} \sum_{\text{training set}} ||\boldsymbol{g}_{\theta_g}(\boldsymbol{X}_p, \boldsymbol{e}_{\theta_e}(\boldsymbol{x}^n))- \boldsymbol{x^n_p}||^2_2.$$
    \item Denote the latent space vectors obtained from the encoder above as $\hat{\boldsymbol{x}}_n = \boldsymbol{e}_{\theta_e} (\bx^n).$ Train stress decoder network $\boldsymbol{h}_{\theta_h}(\bX, \hat{\boldsymbol{x}}_n)$ by $$\min_{\theta_h}\sum_{\text{training set}} ||\boldsymbol{h}_{\theta_h}(\bX_p, \hat{\boldsymbol{x}}_n) - \boldsymbol{\tau}^n_p||^2_2.$$
    \item Train an affine momentum network $\boldsymbol{l}_{\theta_l}(\bX, \xhatn)$ by $$\min_{\theta_l}\sum_{\text{training set}} ||\boldsymbol{l}_{\theta_l}(\bX_p, \hat{\boldsymbol{x}}_n) - \boldsymbol{C}^n_p||^2_2.$$
\end{enumerate}
Here $\theta_{*}$ denotes network weights. Additional training details and network architecture are listed in the supplementary material.

If the problem parameter $\mu$ contains information about return mapping, we can make the stress decoder explicitly depend on $\mu$, i.e., $\boldsymbol{h}(\bX, \hat{\boldsymbol{x}}, \mu).$

Together with the three neural networks, we have equipped ourselves with all ingredients needed to perform one step of MPM algorithm.

\section{reduced-order model: dynamics}
\label{sec:reduce:dynamics}
\begin{figure}[t]
  \centering
  \includegraphics[width=\linewidth]{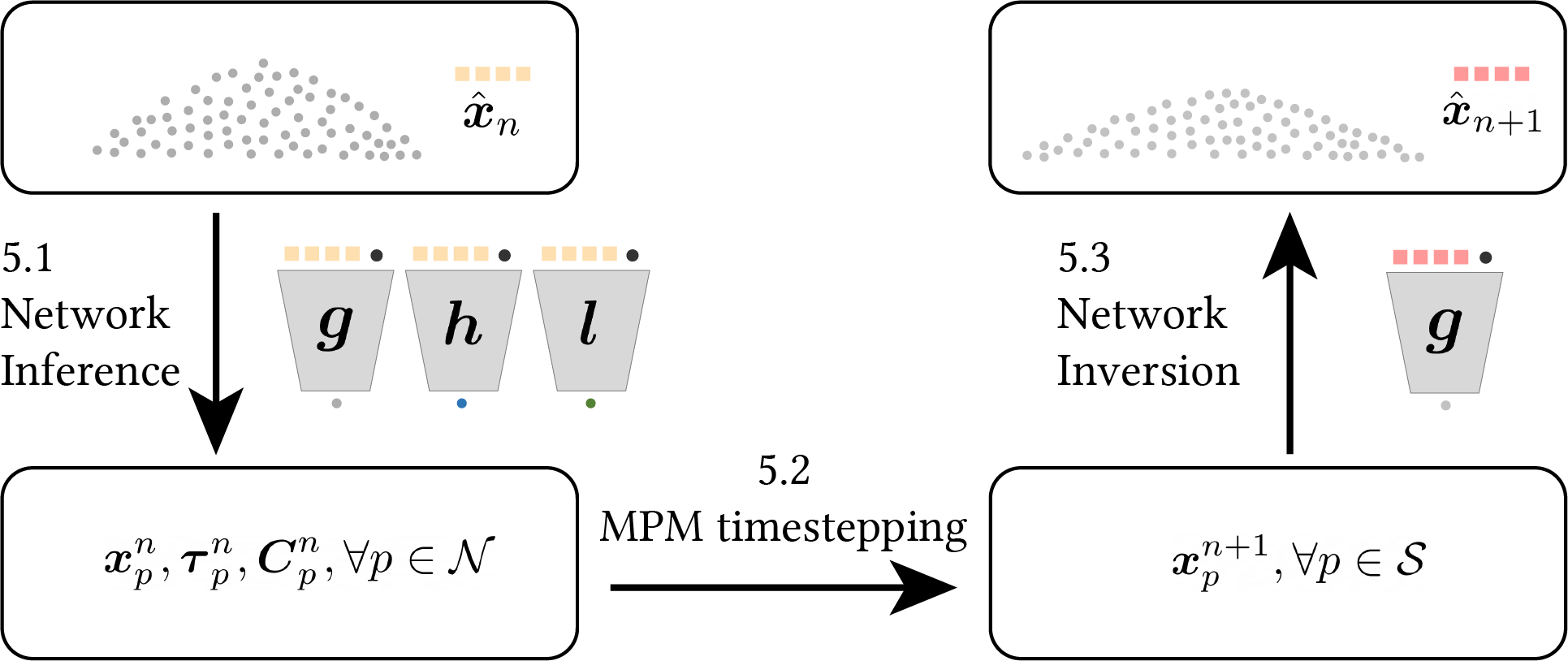}
  \caption{Latent space dynamics. We time-step the latent space via three steps. Each step involves a small spatial subset $\mathcal{S}\subset\mathcal{N}\subset \mathcal{P}$ of the original full-order MPM particles.}
  \Description{}
  \label{fig:latent_dynamics}
\end{figure}

After training, we can run new simulations by time-stepping in the latent space $\mathcal{L}$, from $\xhat_n$ to $\xhat_{n+1}.$ For this, we follow the projection-based ROM approach by \citet{chen2023model}. Our projection-based ROM approach takes three steps: (1) network inference, (2) MPM time-stepping, and (3) network inversion. The pipeline is shown in \cref{fig:latent_dynamics}. As we will see, since the dimension of the manifold $r$ is much much smaller than that of the full order problem $|\mathcal{P}|,$ only a small subset $\mathcal{S}\subset \mathcal{P}$ of particles, which are named sample particles, are needed to determine $\xhat$ dynamics. Nevertheless, due to the non-local nature of MPM, time integration of this subset will involve a larger subset $\mathcal{N}\subset\mathcal{P}$, which we refer to as integration particles. Note that $\mathcal{S}\subset\mathcal{N}$ and $r \leq 3 |\mathcal{S}| < 3|\mathcal{N}| \ll|\mathcal{P}|.$ These sample and integration particles bear similarities to the cubature points often employed in reduced-order FEM \citep{an2008optimizing}. Their exact choice will be deferred to \cref{sec:construct_neighbors}.

 \subsection{Network inference}
 At timestep $t_n,$ given $\xhat_n,$ the states for all initial location $\bX\in\Omega_0,$ and in particular for the integration particles $p \in \mathcal{N}$ with initial position $\bX_p$ can be obtained by inferencing the neural networks $$\bx_p^n = \boldsymbol{g}(\bX_p, \xhat_n), \quad \boldsymbol{v}_p^n = \frac{1}{\Delta t} (\boldsymbol{g}(\bX_p, \xhat_n) - \boldsymbol{g}(\bX_p, \xhat_{n-1})),$$
 $$\boldsymbol{\tau}_p^n = \boldsymbol{h}(\bX_p, \xhat_n), \quad \boldsymbol{C}_p^n = \boldsymbol{l}(\bX_p, \xhat_n).$$
 Note that the particle velocity here is obtained by backward differencing the position field, consistent with the explicit MPM framework.
 \subsection{MPM time-stepping}
 One step of the MPM algorithm (\cref{sec:mpm algo}) is performed on the integration particles $\mathcal{N}$ to advance to $t_{n+1}.$ Integrating all the particles belonging to $\mathcal{N}$ guarantees that the states on sample particles $\bx_p^{n+1} |_{p \in \mathcal{S}}$ are the same as if we perform the full-order MPM on all particles $p \in \mathcal{P}$. There is no approximation in this step.
\subsection{Network Inversion}
 With the new particle positions at $t_{n+1}$ in hand, we are able to find the corresponding $\xhat_{n+1}$ by inverting the neural deformation field,
\begin{equation}
    \xhat_{n+1} = \argmin_{\xhat \in \mathbb{R}^r} \sum_{p \in \mathcal{S}} || \boldsymbol{g}(\bX_p, \xhat) - \bx_p^{n+1}||^2_2.
    \label{eqn:network_inversion}
\end{equation}
In this optimization problem, both the unknown $\xhat_{n+1}$ and the number of summands $|\mathcal{S}|$ are significantly reduced. As the latent space trajectory generally evolves smoothly, with $\xhat_n$ as an initial guess, \cref{eqn:network_inversion} can be rapidly solved via the Gauss-Newton method \citep{nocedal1999numerical}, converging in 2-3 iterations. We can optionally further speed up this nonlinear solver via a first-order Taylor approximation \citep{chen2023model}.

\subsection{Construction of Sample and Integration Particles}
\label{sec:construct_neighbors}
The least-squares problem is well-posed provided $|\mathcal{S}| \geq r/3.$ The projection will be more accurate if a decent number of sample particles can reflect the deformation of the geometry. For example, there should not be a group of sample particles that stand still in a corner. Moreover, the sample particles can be different (in terms of both quantities and spatial distributions) at different time steps. For simplicity, we fix a set of sample particles throughout $[0, T].$ Currently, we choose sample particles via either user-defined heuristics (\cref{sec:fracture} cake cutting) or random sampling (see all other experiments). Future work may consider further optimizing the sample particle choices \citep{an2008optimizing}. 

\begin{figure}[t]
  \centering
  \includegraphics[width=\linewidth]{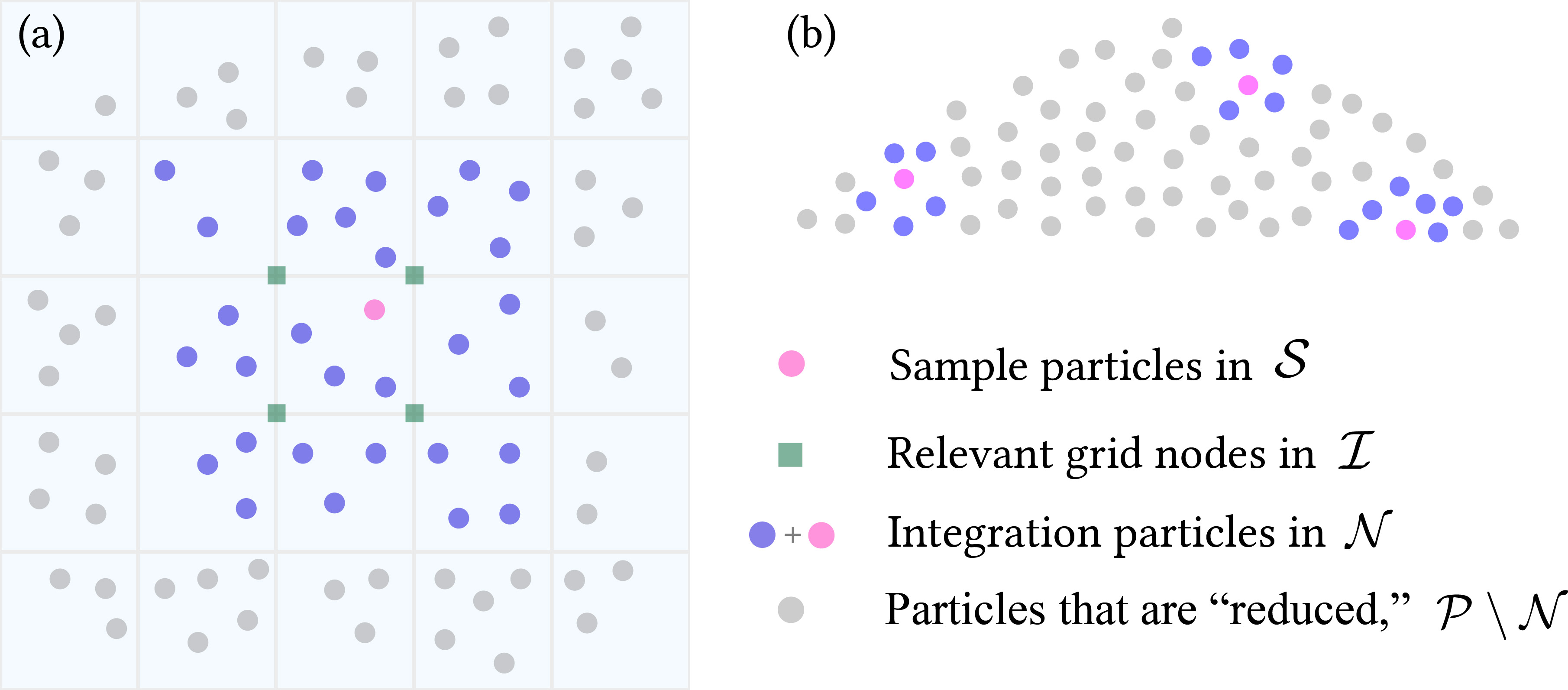}
  \caption{(a) Construction of sample particles and integration particles; (b) Sample and integration particles across the domain.}
  \Description{}
  \label{fig:neighbor}
\end{figure}

Once $\mathcal{S}$ is chosen, we assemble a group of integration particles containing just enough information to evolve sample particles to the next time step. This is done by the following: a. identify the set of grid nodes relevant to $\mathcal{S}$ as $\mathcal{I} = \{ \text{all grid nodes } i: \exists p \in \mathcal{S} \text{ s.t. } N_i(\bx_p)\neq 0 \},$ b. identify the set of integration particles relevant to $\mathcal{I}$ as $\mathcal{N} = \{ \text{all particles } p \in \mathcal{P} : \exists i \in \mathcal{I} \text { s.t. } N_i\left(\boldsymbol{g}(\boldsymbol{X}_p ; \hat{\boldsymbol{x}}_n)\right.\neq 0\}.$ An illustration is shown in \cref{fig:neighbor}.
\section{Experiments}
\label{sec:experiments}
We validate the proposed reduced-order framework on a wide range of elastoplastic examples. The choice of the problem parameter $\mu$ is stated in each experiment. The Experimental statistics are summarized in \cref{table:simulation}. 

In addition to visual results, we will also report the total relative deformation error across space and time, \begin{equation}
\delta = \sqrt{\frac{\sum_{n=1,2,..,N, p
\in \mathcal{P}}  ||\boldsymbol{g}(\bX_p,\xhat_n)-\boldsymbol{\phi}(\bX_p, t_n)||^2 }{\sum_{n=1,2,..,N, p
\in \mathcal{P}}  ||\boldsymbol{\phi}(\bX_p, t_n)||^2 }}. \label{eqn:error def}\end{equation}
Throughout this section, dataset $\mathcal{D}$ is always split as non-overlapping $\mathcal{D}_{\text{train}}$ and $\mathcal{D}_{\text{test}}.$ Neural fields are constructed with $\mathcal{D}_{\text{train}}$ and validated on $\mathcal{D}_{\text{test}}.$  Furthermore, we will report the dimension reduction ratio defined by $\gamma = 3|\mathcal{P}| / r$, i.e., the dimension of the full-order model divided by the latent space dimension. See the supplementary material for additional details regarding experiments, the training dataset, generalizability, extrapolation, and elastoplastic models.

\subsection{Fracture}
\label{sec:fracture}
\begin{figure}
  \centering
  \includegraphics[width=\linewidth]{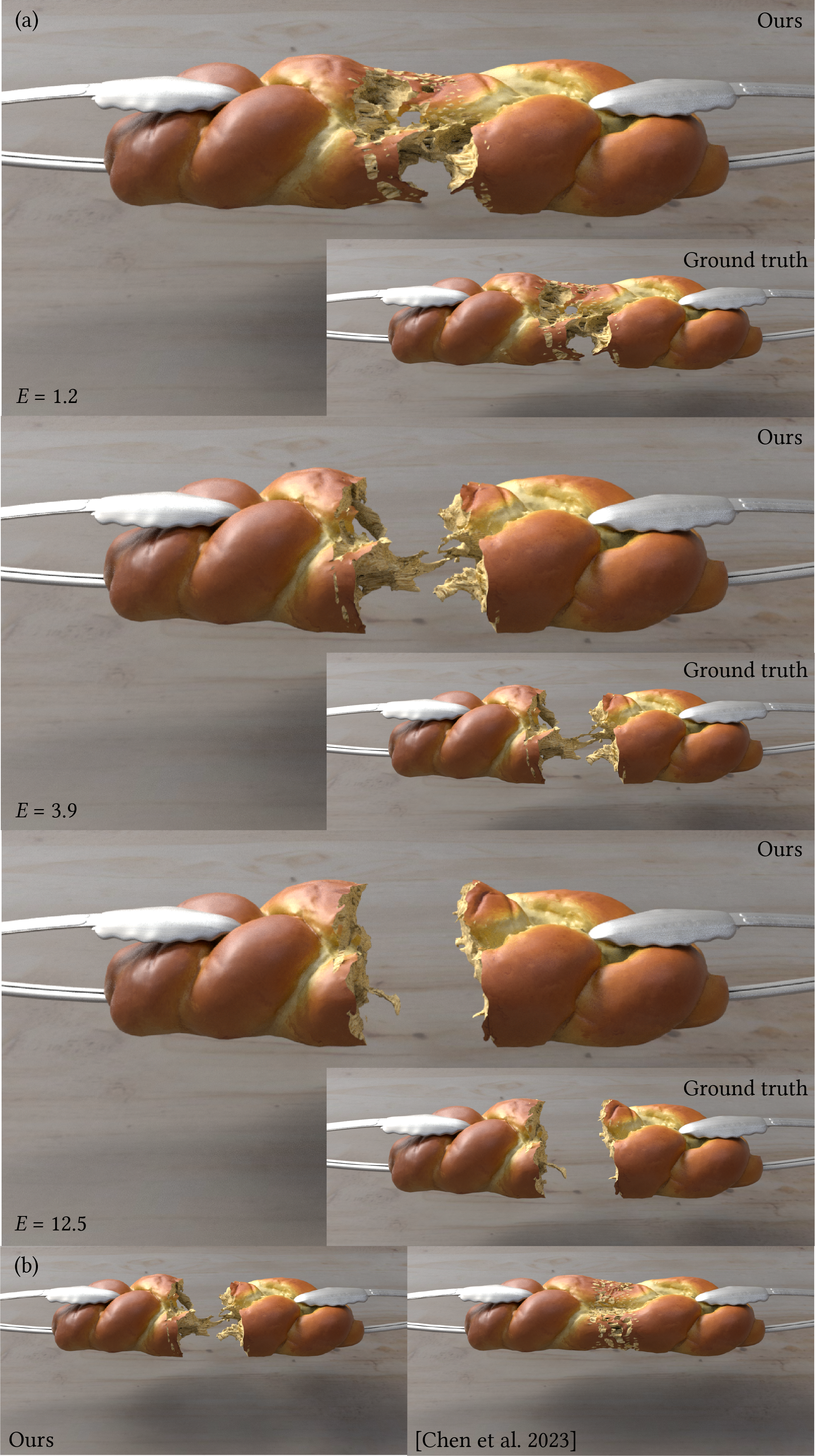}
  \caption{Tear a piece of bread. Our method accurately captures the tearing behavior at different elastic moduli. Due to a lack of accurate stress representation and the inaccurate deformation gradients computed from neural fields, the baseline approach by \citep{chen2023model} fails to capture the fracturing behavior.}
  \Description{The top right subplots show the corresponding ground truth.}
  \label{fig:bread_compare}
\end{figure}
One remarkable feature of our neural stress field is its ability to capture fracture. We first simulate the tearing of a piece of bread with $|\mathcal{P}| = 4 \times 10^4$ particles governed by pure elasticity under different Young's moduli. The problem parameter $\mu$ is the Young's modulus of the material. Weak elements are inserted in the middle region to seed the fracture. \Cref{fig:bread_compare}(a) shows that our method is able to accurately generate the fracture pattern under various unseen Young's moduli. In MPM, numerical fracture happens when two (or more) sets of particles cannot see each other via the grid, after which the deformation gradient $\boldsymbol{F}$ for the two (or more) fractured pieces evolve independently. Thus, in this scenario, $\boldsymbol{F}$ computed from $\frac{\partial \boldsymbol{\phi}}{\partial \bX}$ or its approximation $\frac{\partial \boldsymbol{f}}{\partial \bX}$ would provide a misleading stress $\boldsymbol{\tau}$ that is \textit{not} what is being used in the MPM setting. As is shown in \cref{fig:bread_compare}(b), the baseline method by \citet{chen2023model}, which uses $\boldsymbol{\tau} \approx \boldsymbol{\tau}(\frac{\partial \boldsymbol{f}}{\partial \bX})$, fails to reconstruct a clean fracture. Our neural stress field, on the other hand, explicitly equips the reduced-order model with the \textit{bona fide} stress $\boldsymbol{\tau}$ that \textit{is} used in the ground truth MPM simulation. Here $\mathcal{S}$ consists of $450$ randomly chosen particles $p\in \mathcal{P},$ and the total relative deformation error is $\delta = 1.2\%,$ as is defined in \cref{eqn:error def}. The error for the baseline method is $\delta = 6.5\%.$
\begin{figure}
  \centering
  \includegraphics[width=\linewidth]{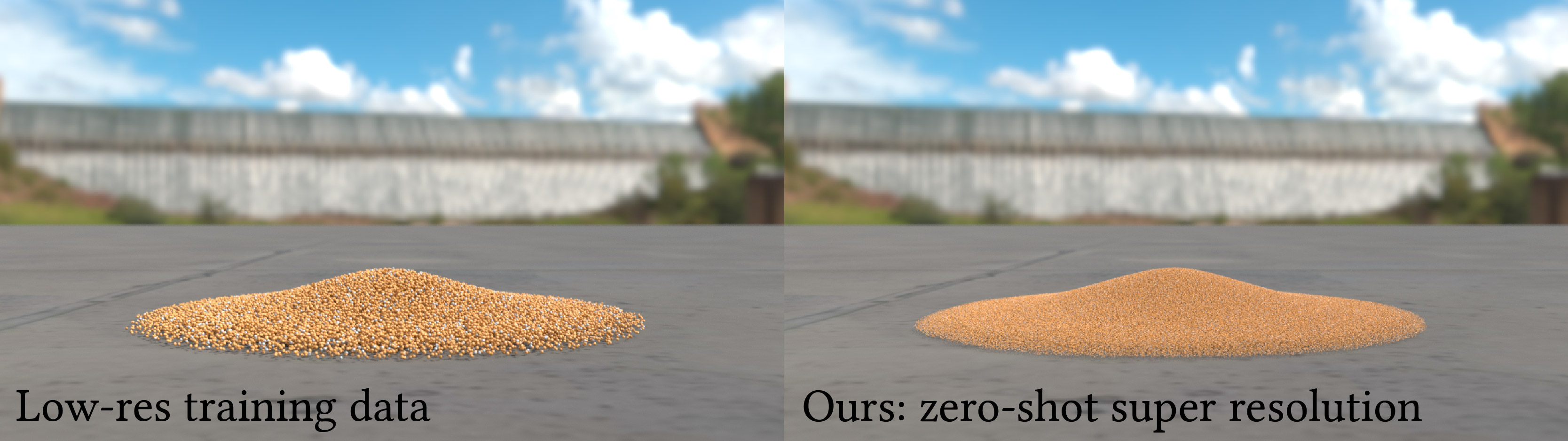}
  \caption{After training on low-res simulation (left), our method can directly infer high-resolution results (right) by querying the continuous neural deformation field. No additional post-processing is needed.}
  \label{fig:sand_highres}
\end{figure}

Our neural stress field is also applicable to fracture with plastic models, such as von Mises plasticity, as is shown in cake cutting in \cref{fig:cake_placeholder}. Here we adopt the plasticity model in \citep{wang2020material}. The cake is simulated with $|\mathcal{P}|= 2\times 10^5$ particles. A spatula is slicing the cake at different angles, represented by the problem parameter $\mu$. We select $700$ particles clustered toward the middle and then reduce the sample size to $400$ after $\frac{T}{2}.$ The number of integration particles is $1.35 \times 10^4$ on average. The full-order and reduced-order MPM simulators are both implemented in WARP \cite{warp2022} under double precision. The neural networks are implemented in PyTorch. The total wall clock time of the full-order simulation is $14.495s,$ while the wall time of our reduced method is $1.417s.$ We achieve an overall speedup of $10.23\times$ with an error of $1.3\%.$  In general, since the dynamics are constrained to the low-dimensional manifold, we are also able to take a larger time step ($1.5\Delta t.$) at deployment time. In both fracture examples we choose $r=6,$ and $\gamma=2 \times 10^4$ and $1 \times 10^6,$ respectively. With our reduced model, we are also able to achieve considerable memory saving. In this scenario, the average memory consumption of the full-order MPM model is 1.61G, while ours is 0.79G, including both latent space physics and neural networks. The computing setup is detailed in the supplementary material.

\subsection{Sand plasticity}
\begin{figure}
  \centering
  \includegraphics[width=\linewidth]{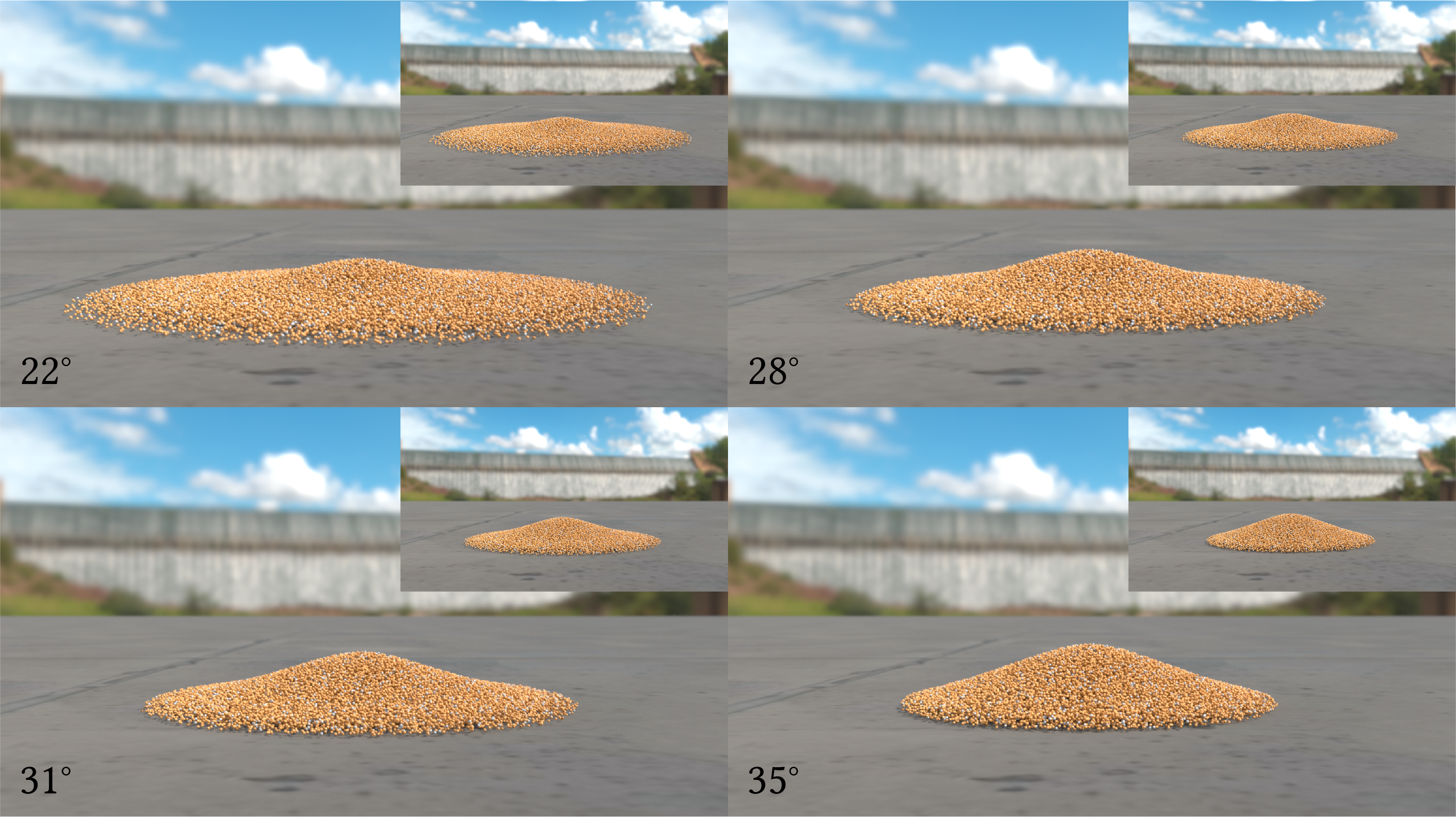}
  \caption{Simulate column collapse for sand under varying friction angles.}
  \label{fig:sand_compare}
\end{figure}
MPM is particularly suitable for simulating granular media. We simulate a column of sand falling onto the ground under gravity. Here $\mu$ represents different friction angles.  Our neural stress field can perfectly capture such a noisy stress distribution and yields excellent results on $\mathcal{D}_{\text{test}},$ with an average error of $\delta = 0.4\%.$ (See \cref{fig:sand_compare}) The ground truth is simulated with $|\mathcal{P}|=72,000$ particles, while we set $r=6, \gamma = 3.6 \times 10^4$ and $|\mathcal{S}|=150.$ The memory consumption of full-order MPM for this scenario is 0.91G, and that of our reduced model is 0.65G.

Once trained using a low-resolution simulation, our approach can arbitrarily boost the resolution with no cost by simply evaluating the neural deformation fields at more $\bX_p \in \Omega_0$. In \cref{fig:sand_highres}, we boost the resolution by 100$\times$ when running latent space dynamics. 
\begin{table*}[t]
    \centering
    \fontsize{.75em}{.5em}\selectfont
    \caption{\textbf{Simulation and reduction statistics.}}
    \begin{tabular}{l@{\hskip1.5pt}lrrrrrrrrrrrrrr}
        \toprule
        \multicolumn{1}{l}{Scene} & Figure & Model & $\Delta t$ & $\Delta x$ & \# of particles & Elasticity/Plasticity & $\dim(\mathcal{L})$ & MLP size for $\boldsymbol{g}, \boldsymbol{h},$ and $\boldsymbol{l},$ respectively &Error \\ \midrule
        Bread & \ref{fig:bread_compare} & Fixed corotated elasticity & 0.001 & 0.0063 & 40,000 & $E \in [1.0,13.0]$ & $r=6$ & $(5, 48\cdot 3), (5, 64\cdot 6), (5, 64\cdot 9)$ &1.2\% \\\midrule
        Cake & \ref{fig:cake_placeholder} & von Mises with softening & 0.0016 & 0.0063 & 200,000 & $\tau_y = 0.1, \theta = 0.03$ & $r=6$ & $(5, 48\cdot 3), (5, 64\cdot 6), (5, 64\cdot 9)$ &1.3\%\\\midrule
        Sand & \ref{fig:sand_compare} \ref{fig:sand_highres}  & Drucker-Prager & 0.002 & 0.0067 & 71,363 & $\phi_f \in [20^{\circ}, 40^{\circ}]$& $r=6$  & $(5, 48 \cdot 3), (5, 72\cdot 6), (5, 72\cdot 9)$ &0.4\% \\\midrule
        Metal & \ref{fig:hardening_placeholder} \ref{fig:hardening_placeholder2}  & von Mises with hardening & 0.0015 & 0.01 & 49,978 & $\tau_Y = 0.05, \xi \in [0.0, 0.2]$ & $r=5$& $(5, 32 \cdot 3), (5, 48\cdot 6), (5, 48\cdot 9)$ &0.2-0.5\% \\\midrule
        Toothpaste & \ref{fig:toothpaste_placeholder1} \ref{fig:toothpaste_placeholder2} & Herschel-Bulkley & 0.001 & 0.0063 & 21,811 & $\tau_Y = 0.05, \eta=0.17$ & $r=6$& $(5, 32 \cdot 3), (5, 48\cdot 6), (5, 48\cdot 9)$ &0.6-1.8\% \\\midrule
        Jelly cube & \ref{fig:baseline1}  & Fixed corotated elasticity& 0.01 & 0.02 & 10,000 & $E = 1.0$ & $r=5$ & $(5, 32 \cdot 3), (5, 64\cdot 6), (5, 64\cdot 9)$&0.2\% \\\midrule
        Squishy ball & \ref{fig:baseline2}  & Fixed corotated elasticity & 0.002 & 0.0067 & 97,857 & $E=40.0$ & $r=6$ & $(5, 48 \cdot 3), (5, 64 \cdot 6), (5, 72\cdot 9)$ &0.2\%\\\midrule
        
    \end{tabular}
    \label{table:simulation}
\end{table*}

\subsection{Metal plasticity}
\begin{figure}[t]
  \centering
  \includegraphics[width=\linewidth]{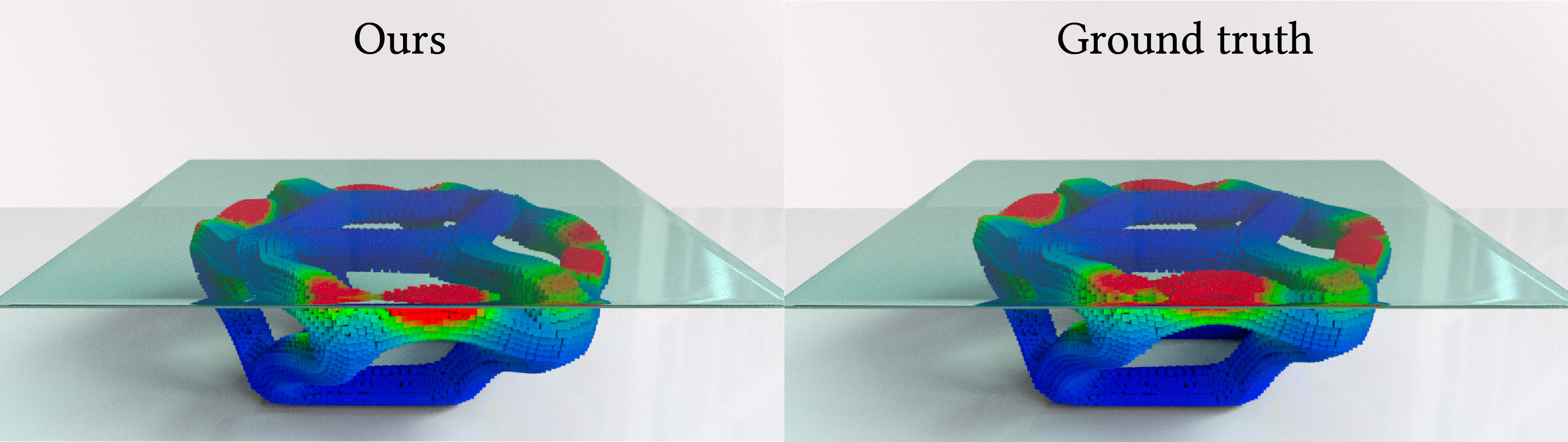}
  \caption{Unlike end-to-end ML frameworks that can only predict particle positions, our first-principal-based reduced-order approach also matches stress quantitatively.}
  \label{fig:hardening_placeholder2}
\end{figure}

Our neural stress field can also handle history-based plasticity models, such as the effect of hardening \citep{wang2019simulation}. The squeezing and bouncing back of a metal frame is simulated with von Mises return mapping under different hardening coefficients $\mu=\tau_Y.$ In the ground truth simulation, the yield condition is $y(\boldsymbol{\tau})<0$, and thus the return mapping is constantly updated to account for hardening, de facto making the yield condition another path-based state $y(\cdot) = y(\cdot, t).$ Since our neural stress field directly approximates the stress computed \textit{after} the return mapping, such complexity is circumvented. In other words, the hardening state is implicitly learned by our neural stress field $\boldsymbol{h}.$ 
 \begin{figure}
  \centering
  \includegraphics[width=\linewidth]{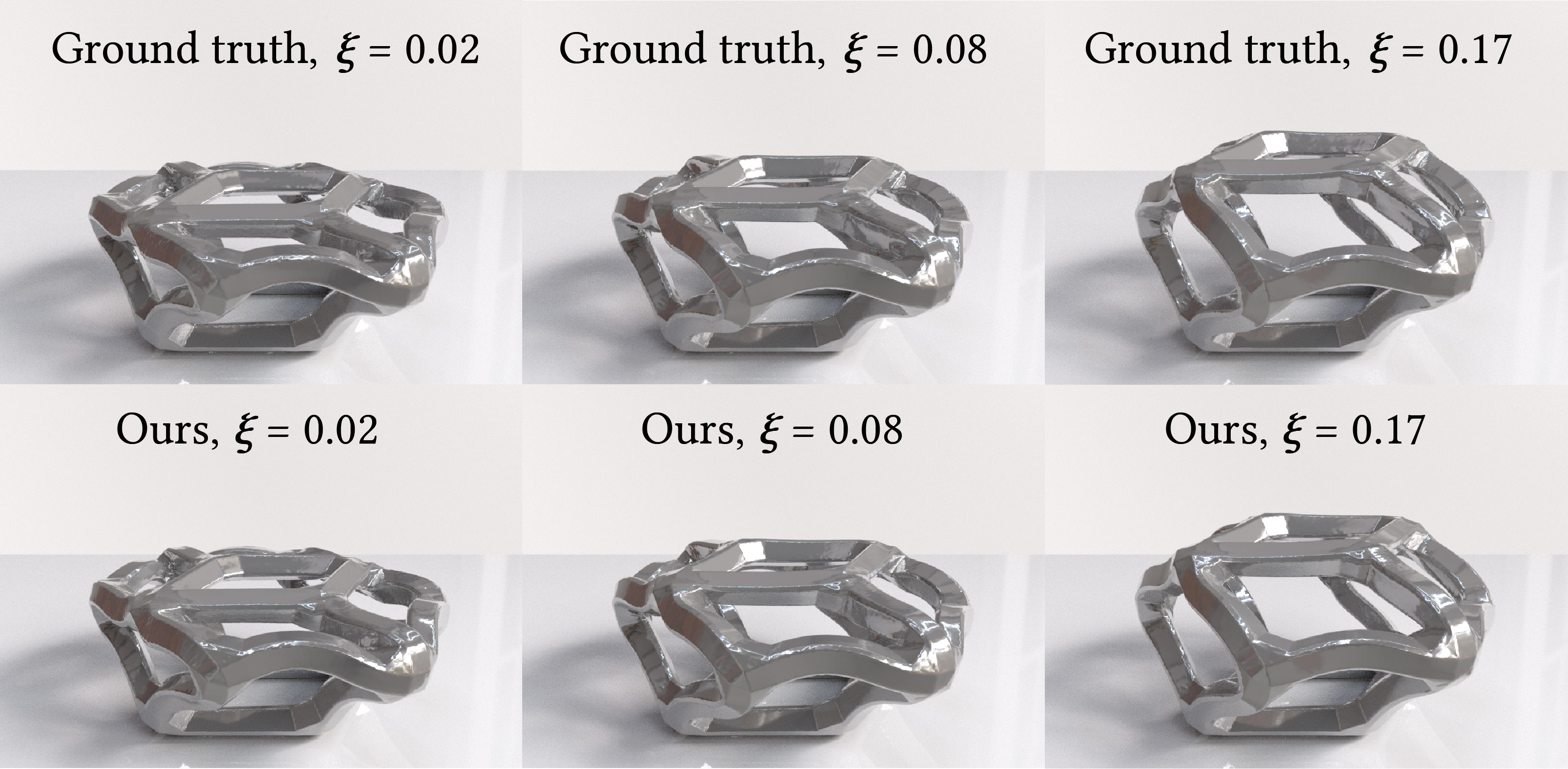}
  \caption{Our neural stress field can capture the hardening effect under different hardening coefficients.}
  \label{fig:hardening_placeholder}
\end{figure}
In \cref{fig:hardening_placeholder}, we compare our deployment results and ground truth under different hardening coefficients.A sampling of $|\mathcal{S}| = 50$ particles out of $|\mathcal{P}| = 49,978$ yields a remarkably small error of $\delta = 0.2 \%$ averaging over all testing data, where we choose $r=5$ and $\gamma = 29.987.$ While end-to-end ML frameworks \citep{sanchez2020learning} can only predict particle positions at rollout time, our PDE-based reduced-order model captures various physical quantities beyond positions. Indeed, our neural stress field can also accurately predict the stress distribution, as is shown in \cref{fig:hardening_placeholder2}
Furthermore, we can sample even fewer points to still obtain reasonably good results. Sampling only $20$ particles results in an error of $\delta = 0.5\%,$ while sampling merely $30$ particles results in an error of $\delta = 0.3\%,$ and the results are almost indistinguishable visually compared with the ground truth. In addition, with a randomly chosen $30$ sample particles, and with the timestep in deployment set to $1.5\Delta t,$ we are able to speed up the total wall clock time from $7.83s$ in the full-order MPM to $1.55s$ in the reduced model, achieving a speedup more than $5\times.$ In this setup, the full-order MPM memory consumption is 0.82G, while ours is 0.51G.

\subsection{Non-Newtonian fluids}
\begin{figure}
  \centering
  \includegraphics[width=\linewidth]{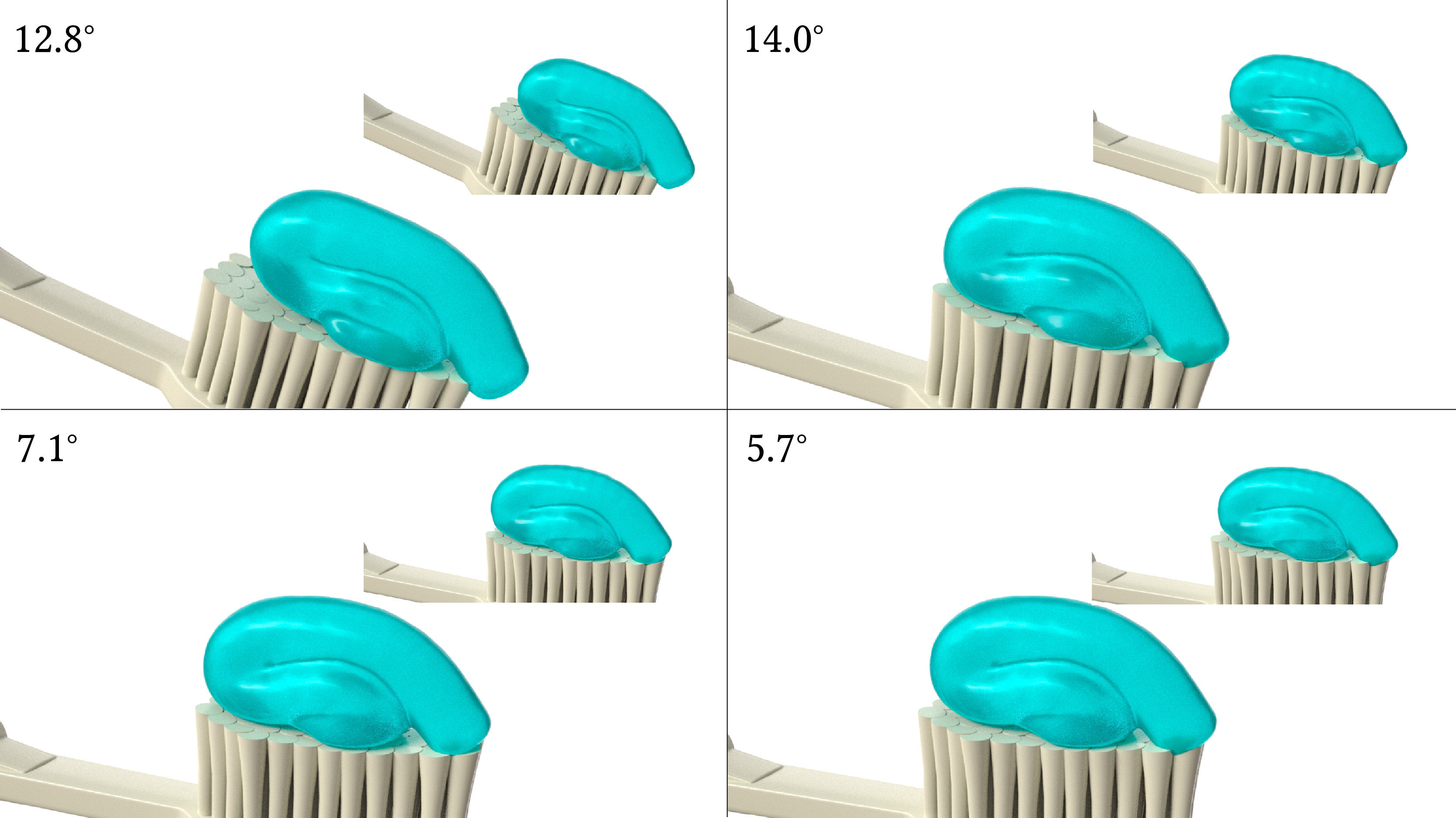}
  \caption{A ribbon of toothpaste is smeared onto a toothbrush held at different angles. The four subplots show our deployment results with $50$ sample points. The corresponding ground truth is shown in the top right corner of each subplot.}
  \label{fig:toothpaste_placeholder1}
\end{figure}
\begin{figure}[t]
  \centering
  \includegraphics[width=\linewidth]{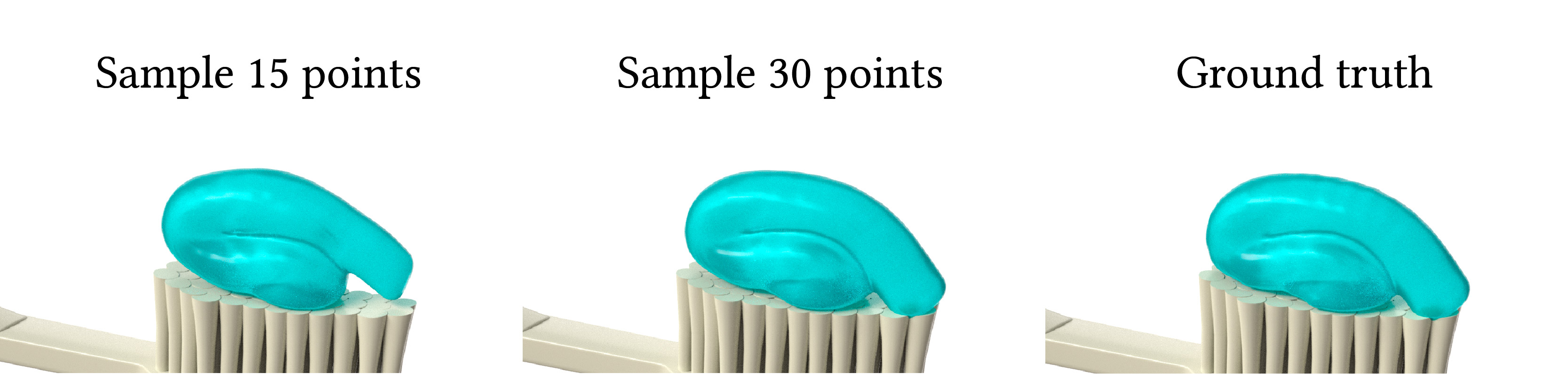}
  \caption{Results for different numbers of sampling points are shown. For this example where the toothpaste is held at $7.1^{\circ},$ sampling $15$ randomly chosen particles yields an error of $\delta = 1.8\%,$ while sampling $30$ yields an error of $\delta = 0.9\%.$}
  \label{fig:toothpaste_placeholder2}
\end{figure}

We simulate a ribbon of toothpaste smeared onto a toothbrush holding at different angles with $|\mathcal{P}| = 45,412$ particles (See \cref{fig:toothpaste_placeholder1}). Here, the problem parameter represents different boundary conditions, i.e., toothbrush inclination. We choose $r=6,$ and thus $\gamma = 22,706.$ We follow the Herschel-Bulkley model in \citep{yue2015continuum}. With just $50$ sampling points, we can predict the dynamics of toothpaste with an averaging total relative deformation error $\delta = 0.6\%.$ Further, the sample size can be even reduced without too much discount on the overall visual quality. 
As is shown in \cref{fig:toothpaste_placeholder2}, with only $30$ points, the deployment result still looks reasonably good, with an error of $\delta = 0.9\%.$

\subsection{Rotation and Collision}
\begin{figure}[t]
  \centering
  \includegraphics[width=\linewidth]{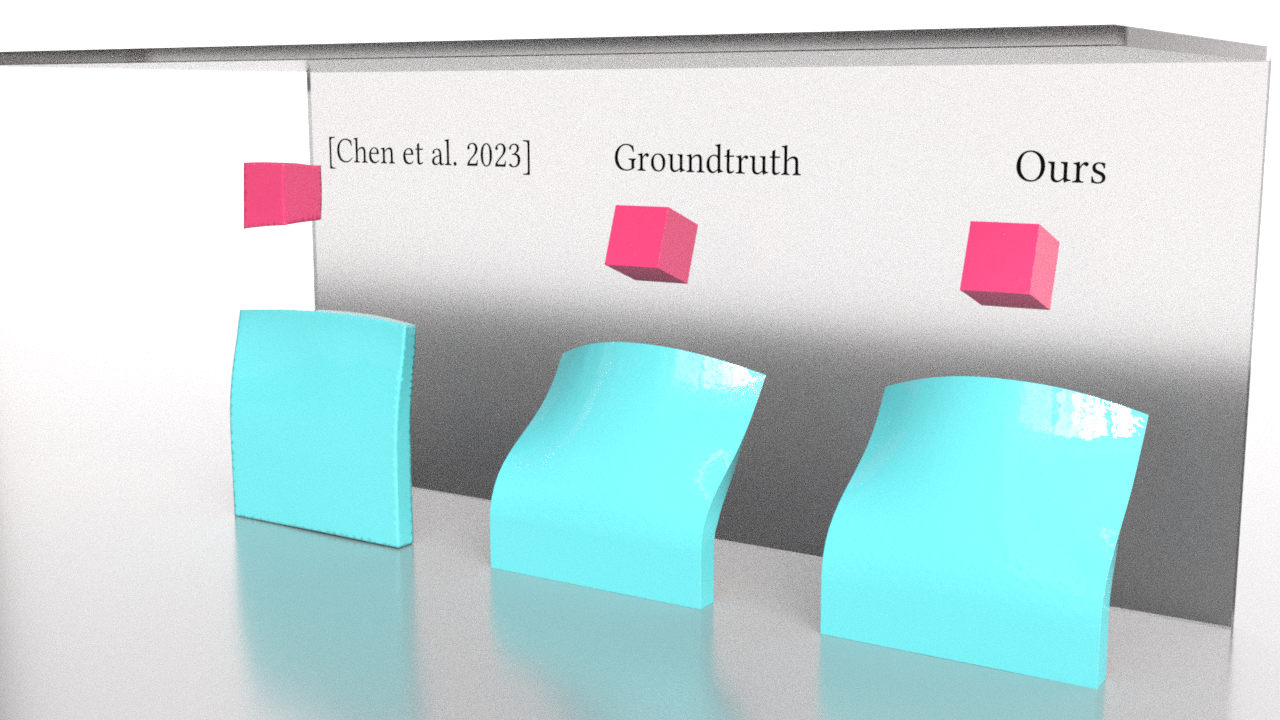}
  \caption{A jelly cube hits onto a jelly wall. Our approach accurately reflects the rotation of the cube. The baseline approach \cite{chen2023crom} is much more dissipative since it does not support angular momentum. The error $\delta$ for our approach is 0.20\%, and for the baseline approach is 16.6\%. The problem parameter $\mu$ represents different initial velocities of the jelly cube.}
  \Description{xxx}
  \label{fig:baseline1}
\end{figure}
We simulate a collision scenario that yields salient rotation (\cref{fig:baseline1}) with $|\mathcal{P}|=10^4$ particles. With a manifold dimension of $r=5$ and $|\mathcal{S}|=50$ sample particles, our approach is able to accurately capture the rotational dynamics. The baseline approach, nevertheless, suffers from noticeable artifacts due to its flawed representation of stress and affine fields.
\begin{figure}[t]
  \centering
  \includegraphics[width=\linewidth]{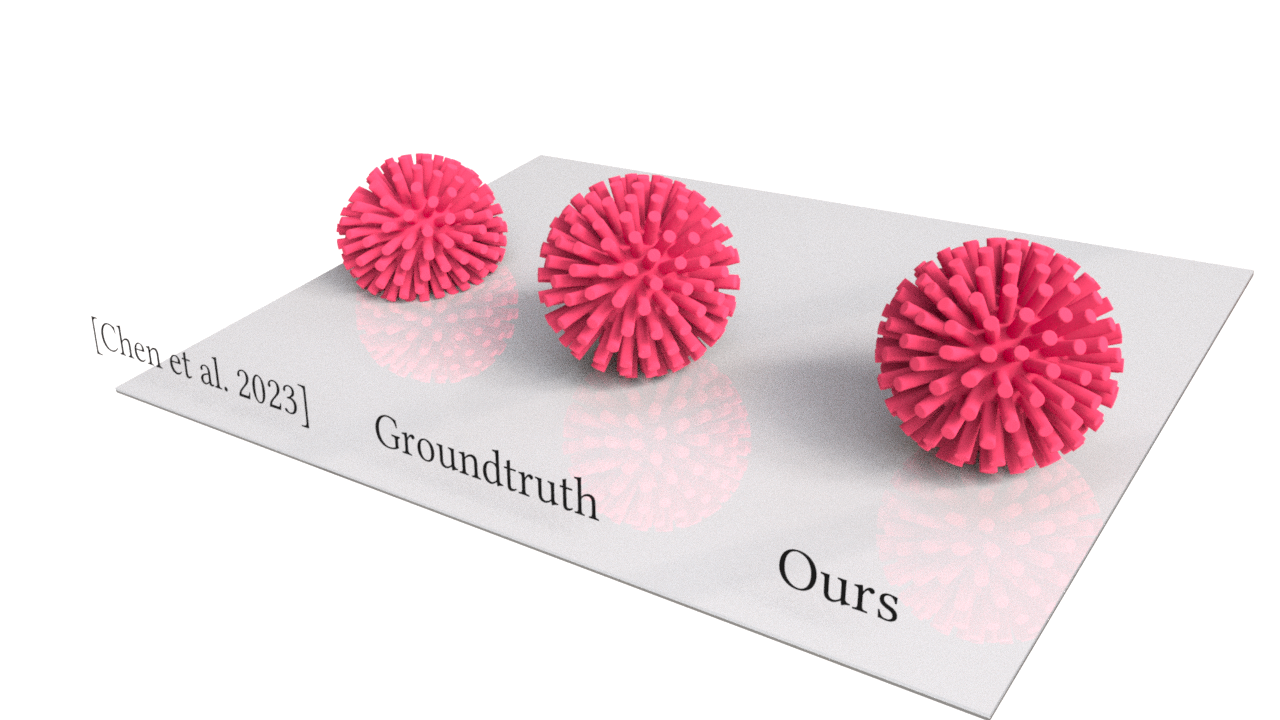}
  \caption{An elastic squishy ball falls onto an inclined plane. Compared with the baseline \cite{chen2023crom}, ours accurately captures both the self-contact and the rotation. The error $\delta$ for our approach is 0.19\%, and for the baseline approach is 4.7\%. $\mu$ represents different inclinations of the plane.}
  \Description{xxx}
  \label{fig:baseline2}
\end{figure}
Our approach can also phonograph complex contact scenarios (\cref{fig:baseline2}). We simulate an elastic squishy ball falling onto an inclined plane with $|\mathcal{P}|=10^5$ particles. The manifold dimension is set to $r=6.$ A randomly selected set of $|\mathcal{S}|=300$ sample particles suffices to delineate the contact of tentacles. The baseline method performs poorly as the affine momentum is missing, and the representation of stress is inaccurate in extreme contact. Notice that we do not need to sample all tentacles to capture their motion; rather, a small $\mathcal{S}$ is used to determine $\xhat$ in the latent space so that our neural deformation and neural stress field can generate their motion. The error $\delta$ for either of the above experiments is less than $0.2\%$. The dimension reduction ratios are $6,000$ and $5 \times 10^4$.

\section{Discussions and future work}
We proposed Neural Stress Fields (NSF), a novel, reduced-order framework for elastoplastic and fracture simulations. NSF significantly alleviates the computational burden of simulating complex elastoplasticity and fracture effects by training a unified, low-dimensional latent space for the neural deformation, stress, and affine fields. Following the training phase, we efficiently conserve computational resources by leveraging these low-dimensional latent variables for evolution. Our approach sets a compelling precedent for multiple potential research trajectories.

\paragraph{Generalization.} Our work supports both interpolation and extrapolation of the training data (see experiments on sand friction angles and bread weak elements). Nevertheless, our approach cannot handle extremely out-of-distribution extrapolation. We trade aggressive generalizability for massive compression and speedup. Future work may consider exploring alternative balancing between generalizability and performance. In addition, for each experiment, we train a network using data from this particular scenario \citep{sifakis2012fem}. An exciting future direction is training on one scenario but generalizing to multiple materials and objects. 

\paragraph{Training time.} Currently, training time is long, between 2hrs and 20hrs. Our target applications are cases where the model would be re-used multiple times. For example, after training, our model can be deployed in VR and gaming applications, where millions of users will interact with it. In these cases, training time is not the main bottleneck. That said, improving training time will help capture larger scenes and accelerate development cycles. 

\paragraph{High-frequency neural fields.} MPM simulations often involve stress fields with high-frequency details and large spatial variations. In practice, we find it challenging to train neural fields that correctly capture these distributions, preventing us from capturing larger scenes. Future work may consider developing more advanced neural architectures \citep{sitzmann2020implicit,tancik2022block} to improve performance when high-frequency details are presented.

\paragraph{Path-dependent plasticity.} Our latent space vector $\xhat_t$ is only determined by the position $\bx_t$. Nevertheless, since plasticity is path-dependent \citep{borja2013plasticity}, the same position field does not imply the same stress field. 
A potential fix to this issue would be to, instead of training two distinct networks, concatenate $\bx$ and $\boldsymbol{\tau}$ and train $\boldsymbol{g}(\bX_p, \boldsymbol{e}([\bx_n, \boldsymbol{\tau}_n])) \approx [\bx^n_p, \boldsymbol{\tau}^n_p].$ In addition, to more explicitly enforce history dependency, future work may consider evolving the latent space according to both stress updates and deformation updates.

\paragraph{Data-free training.} \citet{sharp2023data} introduces a data-free reduced-order modeling framework by incorporating a physics-informed loss term. Extending it to include MPM's plasticity and fracture phenomena is another exciting direction.

\begin{acks}
This work was supported in part by the National Science Foundation (2153851, 2153863, 2023780), Meta, and
Natural Sciences and Engineering Research Council of Canada (Discovery RGPIN-2021-03733). We also express our gratitude to the developers and open-source communities behind PyTorch and NVIDIA Warp.
\end{acks}

\bibliographystyle{ACM-Reference-Format}
\bibliography{reference}


\begin{thebibliography}{68}


\ifx \showCODEN    \undefined \def \showCODEN     #1{\unskip}     \fi
\ifx \showDOI      \undefined \def \showDOI       #1{#1}\fi
\ifx \showISBNx    \undefined \def \showISBNx     #1{\unskip}     \fi
\ifx \showISBNxiii \undefined \def \showISBNxiii  #1{\unskip}     \fi
\ifx \showISSN     \undefined \def \showISSN      #1{\unskip}     \fi
\ifx \showLCCN     \undefined \def \showLCCN      #1{\unskip}     \fi
\ifx \shownote     \undefined \def \shownote      #1{#1}          \fi
\ifx \showarticletitle \undefined \def \showarticletitle #1{#1}   \fi
\ifx \showURL      \undefined \def \showURL       {\relax}        \fi
\providecommand\bibfield[2]{#2}
\providecommand\bibinfo[2]{#2}
\providecommand\natexlab[1]{#1}
\providecommand\showeprint[2][]{arXiv:#2}

\bibitem[Aigerman et~al\mbox{.}(2022)]%
        {aigerman2022neural}
\bibfield{author}{\bibinfo{person}{Noam Aigerman}, \bibinfo{person}{Kunal Gupta}, \bibinfo{person}{Vladimir~G Kim}, \bibinfo{person}{Siddhartha Chaudhuri}, \bibinfo{person}{Jun Saito}, {and} \bibinfo{person}{Thibault Groueix}.} \bibinfo{year}{2022}\natexlab{}.
\newblock \showarticletitle{Neural Jacobian Fields: Learning Intrinsic Mappings of Arbitrary Meshes}.
\newblock \bibinfo{journal}{\emph{arXiv preprint arXiv:2205.02904}} (\bibinfo{year}{2022}).
\newblock


\bibitem[An et~al\mbox{.}(2008)]%
        {an2008optimizing}
\bibfield{author}{\bibinfo{person}{Steven~S An}, \bibinfo{person}{Theodore Kim}, {and} \bibinfo{person}{Doug~L James}.} \bibinfo{year}{2008}\natexlab{}.
\newblock \showarticletitle{Optimizing cubature for efficient integration of subspace deformations}.
\newblock \bibinfo{journal}{\emph{ACM transactions on graphics (TOG)}} \bibinfo{volume}{27}, \bibinfo{number}{5} (\bibinfo{year}{2008}), \bibinfo{pages}{1--10}.
\newblock


\bibitem[Barbi{\v{c}} and James(2005)]%
        {barbivc2005real}
\bibfield{author}{\bibinfo{person}{Jernej Barbi{\v{c}}} {and} \bibinfo{person}{Doug~L James}.} \bibinfo{year}{2005}\natexlab{}.
\newblock \showarticletitle{Real-time subspace integration for {St}. {Venant-Kirchhoff} deformable models}.
\newblock \bibinfo{journal}{\emph{ACM transactions on graphics (TOG)}} \bibinfo{volume}{24}, \bibinfo{number}{3} (\bibinfo{year}{2005}), \bibinfo{pages}{982--990}.
\newblock


\bibitem[Barbi{\v{c}} and Zhao(2011)]%
        {barbivc2011real}
\bibfield{author}{\bibinfo{person}{Jernej Barbi{\v{c}}} {and} \bibinfo{person}{Yili Zhao}.} \bibinfo{year}{2011}\natexlab{}.
\newblock \showarticletitle{Real-time large-deformation substructuring}.
\newblock \bibinfo{journal}{\emph{ACM transactions on graphics (TOG)}} \bibinfo{volume}{30}, \bibinfo{number}{4} (\bibinfo{year}{2011}), \bibinfo{pages}{1--8}.
\newblock


\bibitem[Benner et~al\mbox{.}(2015)]%
        {benner2015survey}
\bibfield{author}{\bibinfo{person}{Peter Benner}, \bibinfo{person}{Serkan Gugercin}, {and} \bibinfo{person}{Karen Willcox}.} \bibinfo{year}{2015}\natexlab{}.
\newblock \showarticletitle{A survey of projection-based model reduction methods for parametric dynamical systems}.
\newblock \bibinfo{journal}{\emph{SIAM review}} \bibinfo{volume}{57}, \bibinfo{number}{4} (\bibinfo{year}{2015}), \bibinfo{pages}{483--531}.
\newblock


\bibitem[Berkooz et~al\mbox{.}(1993)]%
        {berkooz1993proper}
\bibfield{author}{\bibinfo{person}{Gal Berkooz}, \bibinfo{person}{Philip Holmes}, {and} \bibinfo{person}{John~L Lumley}.} \bibinfo{year}{1993}\natexlab{}.
\newblock \showarticletitle{The proper orthogonal decomposition in the analysis of turbulent flows}.
\newblock \bibinfo{journal}{\emph{Annual review of fluid mechanics}} \bibinfo{volume}{25}, \bibinfo{number}{1} (\bibinfo{year}{1993}), \bibinfo{pages}{539--575}.
\newblock


\bibitem[Borja(2013)]%
        {borja2013plasticity}
\bibfield{author}{\bibinfo{person}{Ronaldo~I Borja}.} \bibinfo{year}{2013}\natexlab{}.
\newblock \bibinfo{booktitle}{\emph{Plasticity}}. Vol.~\bibinfo{volume}{2}.
\newblock \bibinfo{publisher}{Springer}.
\newblock


\bibitem[Brackbill and Ruppel(1986)]%
        {brackbill1986flip}
\bibfield{author}{\bibinfo{person}{Jeremiah~U Brackbill} {and} \bibinfo{person}{Hans~M Ruppel}.} \bibinfo{year}{1986}\natexlab{}.
\newblock \showarticletitle{FLIP: A method for adaptively zoned, particle-in-cell calculations of fluid flows in two dimensions}.
\newblock \bibinfo{journal}{\emph{Journal of Computational physics}} \bibinfo{volume}{65}, \bibinfo{number}{2} (\bibinfo{year}{1986}), \bibinfo{pages}{314--343}.
\newblock


\bibitem[Carlberg et~al\mbox{.}(2017)]%
        {carlberg2017galerkin}
\bibfield{author}{\bibinfo{person}{Kevin Carlberg}, \bibinfo{person}{Matthew Barone}, {and} \bibinfo{person}{Harbir Antil}.} \bibinfo{year}{2017}\natexlab{}.
\newblock \showarticletitle{Galerkin v. least-csquares Petrov--Galerkin projection in nonlinear model reduction}.
\newblock \bibinfo{journal}{\emph{J. Comput. Phys.}}  \bibinfo{volume}{330} (\bibinfo{year}{2017}), \bibinfo{pages}{693--734}.
\newblock


\bibitem[Chen et~al\mbox{.}(2022)]%
        {chen2022implicit}
\bibfield{author}{\bibinfo{person}{Honglin Chen}, \bibinfo{person}{Rundi Wu}, \bibinfo{person}{Eitan Grinspun}, \bibinfo{person}{Changxi Zheng}, {and} \bibinfo{person}{Peter~Yichen Chen}.} \bibinfo{year}{2022}\natexlab{}.
\newblock \showarticletitle{Implicit Neural Spatial Representations for Time-dependent PDEs}.
\newblock \bibinfo{journal}{\emph{arXiv preprint arXiv:2210.00124}} (\bibinfo{year}{2022}).
\newblock


\bibitem[Chen et~al\mbox{.}(2021)]%
        {chen2021hybrid}
\bibfield{author}{\bibinfo{person}{Peter~Yichen Chen}, \bibinfo{person}{Maytee Chantharayukhonthorn}, \bibinfo{person}{Yonghao Yue}, \bibinfo{person}{Eitan Grinspun}, {and} \bibinfo{person}{Ken Kamrin}.} \bibinfo{year}{2021}\natexlab{}.
\newblock \showarticletitle{Hybrid discrete-continuum modeling of shear localization in granular media}.
\newblock \bibinfo{journal}{\emph{Journal of the Mechanics and Physics of Solids}}  \bibinfo{volume}{153} (\bibinfo{year}{2021}), \bibinfo{pages}{104404}.
\newblock


\bibitem[Chen et~al\mbox{.}(2023a)]%
        {chen2023model}
\bibfield{author}{\bibinfo{person}{Peter~Yichen Chen}, \bibinfo{person}{Maurizio~M Chiaramonte}, \bibinfo{person}{Eitan Grinspun}, {and} \bibinfo{person}{Kevin Carlberg}.} \bibinfo{year}{2023}\natexlab{a}.
\newblock \showarticletitle{Model reduction for the material point method via an implicit neural representation of the deformation map}.
\newblock \bibinfo{journal}{\emph{J. Comput. Phys.}}  \bibinfo{volume}{478} (\bibinfo{year}{2023}), \bibinfo{pages}{111908}.
\newblock


\bibitem[Chen et~al\mbox{.}(2023b)]%
        {chen2023crom}
\bibfield{author}{\bibinfo{person}{Peter~Yichen Chen}, \bibinfo{person}{Jinxu Xiang}, \bibinfo{person}{Dong~Heon Cho}, \bibinfo{person}{Yue Chang}, \bibinfo{person}{G~A Pershing}, \bibinfo{person}{Henrique~Teles Maia}, \bibinfo{person}{Maurizio~M Chiaramonte}, \bibinfo{person}{Kevin~Thomas Carlberg}, {and} \bibinfo{person}{Eitan Grinspun}.} \bibinfo{year}{2023}\natexlab{b}.
\newblock \showarticletitle{{CROM}: Continuous Reduced-Order Modeling of {PDE}s Using Implicit Neural Representations}. In \bibinfo{booktitle}{\emph{The Eleventh International Conference on Learning Representations}}.
\newblock
\urldef\tempurl%
\url{https://openreview.net/forum?id=FUORz1tG8Og}
\showURL{%
\tempurl}


\bibitem[Chen and Zhang(2019)]%
        {chen2019learning}
\bibfield{author}{\bibinfo{person}{Zhiqin Chen} {and} \bibinfo{person}{Hao Zhang}.} \bibinfo{year}{2019}\natexlab{}.
\newblock \showarticletitle{Learning implicit fields for generative shape modeling}. In \bibinfo{booktitle}{\emph{Proceedings of the IEEE/CVF Conference on Computer Vision and Pattern Recognition}}. \bibinfo{pages}{5939--5948}.
\newblock


\bibitem[Clevert et~al\mbox{.}(2015)]%
        {clevert2015fast}
\bibfield{author}{\bibinfo{person}{Djork-Arn{\'e} Clevert}, \bibinfo{person}{Thomas Unterthiner}, {and} \bibinfo{person}{Sepp Hochreiter}.} \bibinfo{year}{2015}\natexlab{}.
\newblock \showarticletitle{Fast and accurate deep network learning by exponential linear units (elus)}.
\newblock \bibinfo{journal}{\emph{arXiv preprint arXiv:1511.07289}} (\bibinfo{year}{2015}).
\newblock


\bibitem[Daviet and Bertails-Descoubes(2016)]%
        {daviet2016semi}
\bibfield{author}{\bibinfo{person}{Gilles Daviet} {and} \bibinfo{person}{Florence Bertails-Descoubes}.} \bibinfo{year}{2016}\natexlab{}.
\newblock \showarticletitle{A semi-implicit material point method for the continuum simulation of granular materials}.
\newblock \bibinfo{journal}{\emph{ACM Transactions on Graphics (TOG)}} \bibinfo{volume}{35}, \bibinfo{number}{4} (\bibinfo{year}{2016}), \bibinfo{pages}{1--13}.
\newblock


\bibitem[Ding et~al\mbox{.}(2019)]%
        {ding2019thermomechanical}
\bibfield{author}{\bibinfo{person}{Mengyuan Ding}, \bibinfo{person}{Xuchen Han}, \bibinfo{person}{Stephanie Wang}, \bibinfo{person}{Theodore~F Gast}, {and} \bibinfo{person}{Joseph~M Teran}.} \bibinfo{year}{2019}\natexlab{}.
\newblock \showarticletitle{A thermomechanical material point method for baking and cooking}.
\newblock \bibinfo{journal}{\emph{ACM Transactions on Graphics (TOG)}} \bibinfo{volume}{38}, \bibinfo{number}{6} (\bibinfo{year}{2019}), \bibinfo{pages}{1--14}.
\newblock


\bibitem[Falcon(2019)]%
        {pytorchlightning}
\bibfield{author}{\bibinfo{person}{William Falcon}.} \bibinfo{year}{2019}\natexlab{}.
\newblock \bibinfo{title}{PyTorch Lightning}.
\newblock
\newblock


\bibitem[Fang et~al\mbox{.}(2019)]%
        {fang2019silly}
\bibfield{author}{\bibinfo{person}{Yu Fang}, \bibinfo{person}{Minchen Li}, \bibinfo{person}{Ming Gao}, {and} \bibinfo{person}{Chenfanfu Jiang}.} \bibinfo{year}{2019}\natexlab{}.
\newblock \showarticletitle{Silly rubber: an implicit material point method for simulating non-equilibrated viscoelastic and elastoplastic solids}.
\newblock \bibinfo{journal}{\emph{ACM Transactions on Graphics (TOG)}} \bibinfo{volume}{38}, \bibinfo{number}{4} (\bibinfo{year}{2019}), \bibinfo{pages}{1--13}.
\newblock


\bibitem[Fei et~al\mbox{.}(2019)]%
        {fei2019multi}
\bibfield{author}{\bibinfo{person}{Yun Fei}, \bibinfo{person}{Christopher Batty}, \bibinfo{person}{Eitan Grinspun}, {and} \bibinfo{person}{Changxi Zheng}.} \bibinfo{year}{2019}\natexlab{}.
\newblock \showarticletitle{A multi-scale model for coupling strands with shear-dependent liquid}.
\newblock \bibinfo{journal}{\emph{ACM Transactions on Graphics (TOG)}} \bibinfo{volume}{38}, \bibinfo{number}{6} (\bibinfo{year}{2019}), \bibinfo{pages}{1--20}.
\newblock


\bibitem[Fei et~al\mbox{.}(2021)]%
        {fei2021principles}
\bibfield{author}{\bibinfo{person}{Yun Fei}, \bibinfo{person}{Yuhan Huang}, {and} \bibinfo{person}{Ming Gao}.} \bibinfo{year}{2021}\natexlab{}.
\newblock \showarticletitle{Principles towards Real-Time Simulation of Material Point Method on Modern GPUs}.
\newblock \bibinfo{journal}{\emph{arXiv preprint arXiv:2111.00699}} (\bibinfo{year}{2021}).
\newblock


\bibitem[Fulton et~al\mbox{.}(2019)]%
        {fulton2019latent}
\bibfield{author}{\bibinfo{person}{Lawson Fulton}, \bibinfo{person}{Vismay Modi}, \bibinfo{person}{David Duvenaud}, \bibinfo{person}{David~IW Levin}, {and} \bibinfo{person}{Alec Jacobson}.} \bibinfo{year}{2019}\natexlab{}.
\newblock \showarticletitle{Latent-space Dynamics for Reduced Deformable Simulation}. In \bibinfo{booktitle}{\emph{Computer graphics forum}}, Vol.~\bibinfo{volume}{38}. Wiley Online Library, \bibinfo{pages}{379--391}.
\newblock


\bibitem[Gao et~al\mbox{.}(2018)]%
        {gao2018gpu}
\bibfield{author}{\bibinfo{person}{Ming Gao}, \bibinfo{person}{Xinlei Wang}, \bibinfo{person}{Kui Wu}, \bibinfo{person}{Andre Pradhana}, \bibinfo{person}{Eftychios Sifakis}, \bibinfo{person}{Cem Yuksel}, {and} \bibinfo{person}{Chenfanfu Jiang}.} \bibinfo{year}{2018}\natexlab{}.
\newblock \showarticletitle{GPU optimization of material point methods}.
\newblock \bibinfo{journal}{\emph{ACM Transactions on Graphics (TOG)}} \bibinfo{volume}{37}, \bibinfo{number}{6} (\bibinfo{year}{2018}), \bibinfo{pages}{1--12}.
\newblock


\bibitem[Gonzalez and Stuart(2008)]%
        {gonzalez2008first}
\bibfield{author}{\bibinfo{person}{Oscar Gonzalez} {and} \bibinfo{person}{Andrew~M Stuart}.} \bibinfo{year}{2008}\natexlab{}.
\newblock \bibinfo{booktitle}{\emph{A first course in continuum mechanics}}. Vol.~\bibinfo{volume}{42}.
\newblock \bibinfo{publisher}{Cambridge University Press}.
\newblock


\bibitem[Harlow(1962)]%
        {harlow1962particle}
\bibfield{author}{\bibinfo{person}{Francis~H Harlow}.} \bibinfo{year}{1962}\natexlab{}.
\newblock \bibinfo{booktitle}{\emph{The particle-in-cell method for numerical solution of problems in fluid dynamics}}.
\newblock \bibinfo{type}{{T}echnical {R}eport}. \bibinfo{institution}{Los Alamos National Lab.(LANL), Los Alamos, NM (United States)}.
\newblock


\bibitem[Hegemann et~al\mbox{.}(2013)]%
        {hegemann2013level}
\bibfield{author}{\bibinfo{person}{Jan Hegemann}, \bibinfo{person}{Chenfanfu Jiang}, \bibinfo{person}{Craig Schroeder}, {and} \bibinfo{person}{Joseph~M Teran}.} \bibinfo{year}{2013}\natexlab{}.
\newblock \showarticletitle{A level set method for ductile fracture}. In \bibinfo{booktitle}{\emph{Proceedings of the 12th ACM SIGGRAPH/Eurographics Symposium on Computer Animation}}. \bibinfo{pages}{193--201}.
\newblock


\bibitem[Holmes et~al\mbox{.}(2012)]%
        {holmes2012turbulence}
\bibfield{author}{\bibinfo{person}{Philip Holmes}, \bibinfo{person}{John~L Lumley}, \bibinfo{person}{Gahl Berkooz}, {and} \bibinfo{person}{Clarence~W Rowley}.} \bibinfo{year}{2012}\natexlab{}.
\newblock \bibinfo{booktitle}{\emph{Turbulence, coherent structures, dynamical systems and symmetry}}.
\newblock \bibinfo{publisher}{Cambridge university press}.
\newblock


\bibitem[Hu et~al\mbox{.}(2019)]%
        {hu2019taichi}
\bibfield{author}{\bibinfo{person}{Yuanming Hu}, \bibinfo{person}{Tzu-Mao Li}, \bibinfo{person}{Luke Anderson}, \bibinfo{person}{Jonathan Ragan-Kelley}, {and} \bibinfo{person}{Fr{\'e}do Durand}.} \bibinfo{year}{2019}\natexlab{}.
\newblock \showarticletitle{Taichi: a language for high-performance computation on spatially sparse data structures}.
\newblock \bibinfo{journal}{\emph{ACM Transactions on Graphics (TOG)}} \bibinfo{volume}{38}, \bibinfo{number}{6} (\bibinfo{year}{2019}), \bibinfo{pages}{1--16}.
\newblock


\bibitem[Jiang et~al\mbox{.}(2015)]%
        {jiang2015affine}
\bibfield{author}{\bibinfo{person}{Chenfanfu Jiang}, \bibinfo{person}{Craig Schroeder}, \bibinfo{person}{Andrew Selle}, \bibinfo{person}{Joseph Teran}, {and} \bibinfo{person}{Alexey Stomakhin}.} \bibinfo{year}{2015}\natexlab{}.
\newblock \showarticletitle{The affine particle-in-cell method}.
\newblock \bibinfo{journal}{\emph{ACM Transactions on Graphics (TOG)}} \bibinfo{volume}{34}, \bibinfo{number}{4} (\bibinfo{year}{2015}), \bibinfo{pages}{1--10}.
\newblock


\bibitem[Jiang et~al\mbox{.}(2016)]%
        {jiang2016material}
\bibfield{author}{\bibinfo{person}{Chenfanfu Jiang}, \bibinfo{person}{Craig Schroeder}, \bibinfo{person}{Joseph Teran}, \bibinfo{person}{Alexey Stomakhin}, {and} \bibinfo{person}{Andrew Selle}.} \bibinfo{year}{2016}\natexlab{}.
\newblock \showarticletitle{The material point method for simulating continuum materials}.
\newblock In \bibinfo{booktitle}{\emph{ACM SIGGRAPH 2016 Courses}}. \bibinfo{pages}{1--52}.
\newblock


\bibitem[Kim et~al\mbox{.}(2019)]%
        {kim2019deep}
\bibfield{author}{\bibinfo{person}{Byungsoo Kim}, \bibinfo{person}{Vinicius~C Azevedo}, \bibinfo{person}{Nils Thuerey}, \bibinfo{person}{Theodore Kim}, \bibinfo{person}{Markus Gross}, {and} \bibinfo{person}{Barbara Solenthaler}.} \bibinfo{year}{2019}\natexlab{}.
\newblock \showarticletitle{Deep fluids: A generative network for parameterized fluid simulations}. In \bibinfo{booktitle}{\emph{Computer Graphics Forum}}, Vol.~\bibinfo{volume}{38}. Wiley Online Library, \bibinfo{pages}{59--70}.
\newblock


\bibitem[Kim and Delaney(2013)]%
        {kim2013subspace}
\bibfield{author}{\bibinfo{person}{Theodore Kim} {and} \bibinfo{person}{John Delaney}.} \bibinfo{year}{2013}\natexlab{}.
\newblock \showarticletitle{Subspace fluid re-simulation}.
\newblock \bibinfo{journal}{\emph{ACM Transactions on Graphics (TOG)}} \bibinfo{volume}{32}, \bibinfo{number}{4} (\bibinfo{year}{2013}), \bibinfo{pages}{1--9}.
\newblock


\bibitem[Kim and James(2009)]%
        {kim2009skipping}
\bibfield{author}{\bibinfo{person}{Theodore Kim} {and} \bibinfo{person}{Doug~L James}.} \bibinfo{year}{2009}\natexlab{}.
\newblock \showarticletitle{Skipping steps in deformable simulation with online model reduction}.
\newblock In \bibinfo{booktitle}{\emph{ACM SIGGRAPH Asia 2009 papers}}. \bibinfo{pages}{1--9}.
\newblock


\bibitem[Kingma and Ba(2014)]%
        {kingma2014adam}
\bibfield{author}{\bibinfo{person}{Diederik~P Kingma} {and} \bibinfo{person}{Jimmy Ba}.} \bibinfo{year}{2014}\natexlab{}.
\newblock \showarticletitle{Adam: A method for stochastic optimization}.
\newblock \bibinfo{journal}{\emph{arXiv preprint arXiv:1412.6980}} (\bibinfo{year}{2014}).
\newblock


\bibitem[Kl{\'a}r et~al\mbox{.}(2016)]%
        {klar2016drucker}
\bibfield{author}{\bibinfo{person}{Gergely Kl{\'a}r}, \bibinfo{person}{Theodore Gast}, \bibinfo{person}{Andre Pradhana}, \bibinfo{person}{Chuyuan Fu}, \bibinfo{person}{Craig Schroeder}, \bibinfo{person}{Chenfanfu Jiang}, {and} \bibinfo{person}{Joseph Teran}.} \bibinfo{year}{2016}\natexlab{}.
\newblock \showarticletitle{Drucker-prager elastoplasticity for sand animation}.
\newblock \bibinfo{journal}{\emph{ACM Transactions on Graphics (TOG)}} \bibinfo{volume}{35}, \bibinfo{number}{4} (\bibinfo{year}{2016}), \bibinfo{pages}{1--12}.
\newblock


\bibitem[Lee and Carlberg(2020)]%
        {lee2020model}
\bibfield{author}{\bibinfo{person}{Kookjin Lee} {and} \bibinfo{person}{Kevin~T Carlberg}.} \bibinfo{year}{2020}\natexlab{}.
\newblock \showarticletitle{Model reduction of dynamical systems on nonlinear manifolds using deep convolutional autoencoders}.
\newblock \bibinfo{journal}{\emph{J. Comput. Phys.}}  \bibinfo{volume}{404} (\bibinfo{year}{2020}), \bibinfo{pages}{108973}.
\newblock


\bibitem[Lusch et~al\mbox{.}(2018)]%
        {lusch2018deep}
\bibfield{author}{\bibinfo{person}{Bethany Lusch}, \bibinfo{person}{J~Nathan Kutz}, {and} \bibinfo{person}{Steven~L Brunton}.} \bibinfo{year}{2018}\natexlab{}.
\newblock \showarticletitle{Deep learning for universal linear embeddings of nonlinear dynamics}.
\newblock \bibinfo{journal}{\emph{Nature communications}} \bibinfo{volume}{9}, \bibinfo{number}{1} (\bibinfo{year}{2018}), \bibinfo{pages}{4950}.
\newblock


\bibitem[Macklin(2022)]%
        {warp2022}
\bibfield{author}{\bibinfo{person}{Miles Macklin}.} \bibinfo{year}{2022}\natexlab{}.
\newblock \bibinfo{title}{Warp: A High-performance Python Framework for GPU Simulation and Graphics}.
\newblock \bibinfo{howpublished}{\url{https://github.com/nvidia/warp}}.
\newblock
\newblock
\shownote{NVIDIA GPU Technology Conference (GTC)}.


\bibitem[Mescheder et~al\mbox{.}(2019)]%
        {mescheder2019occupancy}
\bibfield{author}{\bibinfo{person}{Lars Mescheder}, \bibinfo{person}{Michael Oechsle}, \bibinfo{person}{Michael Niemeyer}, \bibinfo{person}{Sebastian Nowozin}, {and} \bibinfo{person}{Andreas Geiger}.} \bibinfo{year}{2019}\natexlab{}.
\newblock \showarticletitle{Occupancy networks: Learning 3d reconstruction in function space}. In \bibinfo{booktitle}{\emph{Proceedings of the IEEE/CVF Conference on Computer Vision and Pattern Recognition}}. \bibinfo{pages}{4460--4470}.
\newblock


\bibitem[Mildenhall et~al\mbox{.}(2020)]%
        {mildenhall2020nerf}
\bibfield{author}{\bibinfo{person}{Ben Mildenhall}, \bibinfo{person}{Pratul~P Srinivasan}, \bibinfo{person}{Matthew Tancik}, \bibinfo{person}{Jonathan~T Barron}, \bibinfo{person}{Ravi Ramamoorthi}, {and} \bibinfo{person}{Ren Ng}.} \bibinfo{year}{2020}\natexlab{}.
\newblock \showarticletitle{Nerf: Representing scenes as neural radiance fields for view synthesis}. In \bibinfo{booktitle}{\emph{European conference on computer vision}}. Springer, \bibinfo{pages}{405--421}.
\newblock


\bibitem[Nocedal and Wright(1999)]%
        {nocedal1999numerical}
\bibfield{author}{\bibinfo{person}{Jorge Nocedal} {and} \bibinfo{person}{Stephen~J Wright}.} \bibinfo{year}{1999}\natexlab{}.
\newblock \bibinfo{booktitle}{\emph{Numerical optimization}}.
\newblock \bibinfo{publisher}{Springer}.
\newblock


\bibitem[Pan et~al\mbox{.}(2023)]%
        {pan2022neural}
\bibfield{author}{\bibinfo{person}{Shaowu Pan}, \bibinfo{person}{Steven~L Brunton}, {and} \bibinfo{person}{J~Nathan Kutz}.} \bibinfo{year}{2023}\natexlab{}.
\newblock \showarticletitle{Neural Implicit Flow: a mesh-agnostic dimensionality reduction paradigm of spatio-temporal data}.
\newblock \bibinfo{journal}{\emph{Journal of Machine Learning Research}} \bibinfo{volume}{24}, \bibinfo{number}{41} (\bibinfo{year}{2023}), \bibinfo{pages}{1--60}.
\newblock


\bibitem[Park et~al\mbox{.}(2019)]%
        {park2019deepsdf}
\bibfield{author}{\bibinfo{person}{Jeong~Joon Park}, \bibinfo{person}{Peter Florence}, \bibinfo{person}{Julian Straub}, \bibinfo{person}{Richard Newcombe}, {and} \bibinfo{person}{Steven Lovegrove}.} \bibinfo{year}{2019}\natexlab{}.
\newblock \showarticletitle{Deepsdf: Learning continuous signed distance functions for shape representation}. In \bibinfo{booktitle}{\emph{Proceedings of the IEEE/CVF Conference on Computer Vision and Pattern Recognition}}. \bibinfo{pages}{165--174}.
\newblock


\bibitem[Qiu et~al\mbox{.}(2023)]%
        {qiu2023sparse}
\bibfield{author}{\bibinfo{person}{Yuxing Qiu}, \bibinfo{person}{Samuel~Temple Reeve}, \bibinfo{person}{Minchen Li}, \bibinfo{person}{Yin Yang}, \bibinfo{person}{Stuart~Ryan Slattery}, {and} \bibinfo{person}{Chenfanfu Jiang}.} \bibinfo{year}{2023}\natexlab{}.
\newblock \showarticletitle{A Sparse Distributed Gigascale Resolution Material Point Method}.
\newblock \bibinfo{journal}{\emph{ACM Transactions on Graphics}} \bibinfo{volume}{42}, \bibinfo{number}{2} (\bibinfo{year}{2023}), \bibinfo{pages}{1--21}.
\newblock


\bibitem[Raissi et~al\mbox{.}(2019)]%
        {raissi2019physics}
\bibfield{author}{\bibinfo{person}{Maziar Raissi}, \bibinfo{person}{Paris Perdikaris}, {and} \bibinfo{person}{George~E Karniadakis}.} \bibinfo{year}{2019}\natexlab{}.
\newblock \showarticletitle{Physics-informed neural networks: A deep learning framework for solving forward and inverse problems involving nonlinear partial differential equations}.
\newblock \bibinfo{journal}{\emph{J. Comput. Phys.}}  \bibinfo{volume}{378} (\bibinfo{year}{2019}), \bibinfo{pages}{686--707}.
\newblock


\bibitem[Sanchez-Gonzalez et~al\mbox{.}(2020)]%
        {sanchez2020learning}
\bibfield{author}{\bibinfo{person}{Alvaro Sanchez-Gonzalez}, \bibinfo{person}{Jonathan Godwin}, \bibinfo{person}{Tobias Pfaff}, \bibinfo{person}{Rex Ying}, \bibinfo{person}{Jure Leskovec}, {and} \bibinfo{person}{Peter Battaglia}.} \bibinfo{year}{2020}\natexlab{}.
\newblock \showarticletitle{Learning to simulate complex physics with graph networks}. In \bibinfo{booktitle}{\emph{International conference on machine learning}}. PMLR, \bibinfo{pages}{8459--8468}.
\newblock


\bibitem[Sharp et~al\mbox{.}(2023)]%
        {sharp2023data}
\bibfield{author}{\bibinfo{person}{Nicholas Sharp}, \bibinfo{person}{Cristian Romero}, \bibinfo{person}{Alec Jacobson}, \bibinfo{person}{Etienne Vouga}, \bibinfo{person}{Paul~G Kry}, \bibinfo{person}{David~IW Levin}, {and} \bibinfo{person}{Justin Solomon}.} \bibinfo{year}{2023}\natexlab{}.
\newblock \showarticletitle{Data-Free Learning of Reduced-Order Kinematics}.
\newblock \bibinfo{journal}{\emph{arXiv preprint arXiv:2305.03846}} (\bibinfo{year}{2023}).
\newblock


\bibitem[Shen et~al\mbox{.}(2021)]%
        {shen2021high}
\bibfield{author}{\bibinfo{person}{Siyuan Shen}, \bibinfo{person}{Yin Yang}, \bibinfo{person}{Tianjia Shao}, \bibinfo{person}{He Wang}, \bibinfo{person}{Chenfanfu Jiang}, \bibinfo{person}{Lei Lan}, {and} \bibinfo{person}{Kun Zhou}.} \bibinfo{year}{2021}\natexlab{}.
\newblock \showarticletitle{High-Order Differentiable Autoencoder for Nonlinear Model Reduction}.
\newblock \bibinfo{journal}{\emph{ACM Trans. Graph.}} \bibinfo{volume}{40}, \bibinfo{number}{4}, Article \bibinfo{articleno}{68} (\bibinfo{date}{jul} \bibinfo{year}{2021}), \bibinfo{numpages}{15}~pages.
\newblock
\showISSN{0730-0301}
\urldef\tempurl%
\url{https://doi.org/10.1145/3450626.3459754}
\showDOI{\tempurl}


\bibitem[Sifakis and Barbic(2012)]%
        {sifakis2012fem}
\bibfield{author}{\bibinfo{person}{Eftychios Sifakis} {and} \bibinfo{person}{Jernej Barbic}.} \bibinfo{year}{2012}\natexlab{}.
\newblock \showarticletitle{FEM Simulation of 3D Deformable Solids: A Practitioner's Guide to Theory, Discretization and Model Reduction}. In \bibinfo{booktitle}{\emph{ACM SIGGRAPH 2012 Courses}} (Los Angeles, California) \emph{(\bibinfo{series}{SIGGRAPH '12})}. \bibinfo{publisher}{Association for Computing Machinery}, \bibinfo{address}{New York, NY, USA}, Article \bibinfo{articleno}{20}, \bibinfo{numpages}{50}~pages.
\newblock
\showISBNx{9781450316781}
\urldef\tempurl%
\url{https://doi.org/10.1145/2343483.2343501}
\showDOI{\tempurl}


\bibitem[Sitzmann et~al\mbox{.}(2020)]%
        {sitzmann2020implicit}
\bibfield{author}{\bibinfo{person}{Vincent Sitzmann}, \bibinfo{person}{Julien Martel}, \bibinfo{person}{Alexander Bergman}, \bibinfo{person}{David Lindell}, {and} \bibinfo{person}{Gordon Wetzstein}.} \bibinfo{year}{2020}\natexlab{}.
\newblock \showarticletitle{Implicit neural representations with periodic activation functions}.
\newblock \bibinfo{journal}{\emph{Advances in Neural Information Processing Systems}}  \bibinfo{volume}{33} (\bibinfo{year}{2020}), \bibinfo{pages}{7462--7473}.
\newblock


\bibitem[Stomakhin et~al\mbox{.}(2013)]%
        {stomakhin2013material}
\bibfield{author}{\bibinfo{person}{Alexey Stomakhin}, \bibinfo{person}{Craig Schroeder}, \bibinfo{person}{Lawrence Chai}, \bibinfo{person}{Joseph Teran}, {and} \bibinfo{person}{Andrew Selle}.} \bibinfo{year}{2013}\natexlab{}.
\newblock \showarticletitle{A material point method for snow simulation}.
\newblock \bibinfo{journal}{\emph{ACM Transactions on Graphics (TOG)}} \bibinfo{volume}{32}, \bibinfo{number}{4} (\bibinfo{year}{2013}), \bibinfo{pages}{1--10}.
\newblock


\bibitem[Sulsky et~al\mbox{.}(1995)]%
        {sulsky1995application}
\bibfield{author}{\bibinfo{person}{Deborah Sulsky}, \bibinfo{person}{Shi-Jian Zhou}, {and} \bibinfo{person}{Howard~L Schreyer}.} \bibinfo{year}{1995}\natexlab{}.
\newblock \showarticletitle{Application of a particle-in-cell method to solid mechanics}.
\newblock \bibinfo{journal}{\emph{Computer physics communications}} \bibinfo{volume}{87}, \bibinfo{number}{1-2} (\bibinfo{year}{1995}), \bibinfo{pages}{236--252}.
\newblock


\bibitem[Tancik et~al\mbox{.}(2022)]%
        {tancik2022block}
\bibfield{author}{\bibinfo{person}{Matthew Tancik}, \bibinfo{person}{Vincent Casser}, \bibinfo{person}{Xinchen Yan}, \bibinfo{person}{Sabeek Pradhan}, \bibinfo{person}{Ben Mildenhall}, \bibinfo{person}{Pratul~P Srinivasan}, \bibinfo{person}{Jonathan~T Barron}, {and} \bibinfo{person}{Henrik Kretzschmar}.} \bibinfo{year}{2022}\natexlab{}.
\newblock \showarticletitle{Block-nerf: Scalable large scene neural view synthesis}. In \bibinfo{booktitle}{\emph{Proceedings of the IEEE/CVF Conference on Computer Vision and Pattern Recognition}}. \bibinfo{pages}{8248--8258}.
\newblock


\bibitem[Treuille et~al\mbox{.}(2006)]%
        {treuille2006model}
\bibfield{author}{\bibinfo{person}{Adrien Treuille}, \bibinfo{person}{Andrew Lewis}, {and} \bibinfo{person}{Zoran Popovi{\'c}}.} \bibinfo{year}{2006}\natexlab{}.
\newblock \showarticletitle{Model reduction for real-time fluids}.
\newblock \bibinfo{journal}{\emph{ACM Transactions on Graphics (TOG)}} \bibinfo{volume}{25}, \bibinfo{number}{3} (\bibinfo{year}{2006}), \bibinfo{pages}{826--834}.
\newblock


\bibitem[Wang(2020)]%
        {wang2020material}
\bibfield{author}{\bibinfo{person}{Stephanie Wang}.} \bibinfo{year}{2020}\natexlab{}.
\newblock \bibinfo{booktitle}{\emph{A Material Point Method for Elastoplasticity with Ductile Fracture and Frictional Contact}}.
\newblock \bibinfo{publisher}{University of California, Los Angeles}.
\newblock


\bibitem[Wang et~al\mbox{.}(2019)]%
        {wang2019simulation}
\bibfield{author}{\bibinfo{person}{Stephanie Wang}, \bibinfo{person}{Mengyuan Ding}, \bibinfo{person}{Theodore~F Gast}, \bibinfo{person}{Leyi Zhu}, \bibinfo{person}{Steven Gagniere}, \bibinfo{person}{Chenfanfu Jiang}, {and} \bibinfo{person}{Joseph~M Teran}.} \bibinfo{year}{2019}\natexlab{}.
\newblock \showarticletitle{Simulation and visualization of ductile fracture with the material point method}.
\newblock \bibinfo{journal}{\emph{Proceedings of the ACM on Computer Graphics and Interactive Techniques}} \bibinfo{volume}{2}, \bibinfo{number}{2} (\bibinfo{year}{2019}), \bibinfo{pages}{1--20}.
\newblock


\bibitem[Wang et~al\mbox{.}(2020a)]%
        {wang2020hierarchical}
\bibfield{author}{\bibinfo{person}{Xinlei Wang}, \bibinfo{person}{Minchen Li}, \bibinfo{person}{Yu Fang}, \bibinfo{person}{Xinxin Zhang}, \bibinfo{person}{Ming Gao}, \bibinfo{person}{Min Tang}, \bibinfo{person}{Danny~M Kaufman}, {and} \bibinfo{person}{Chenfanfu Jiang}.} \bibinfo{year}{2020}\natexlab{a}.
\newblock \showarticletitle{Hierarchical optimization time integration for cfl-rate mpm stepping}.
\newblock \bibinfo{journal}{\emph{ACM Transactions on Graphics (TOG)}} \bibinfo{volume}{39}, \bibinfo{number}{3} (\bibinfo{year}{2020}), \bibinfo{pages}{1--16}.
\newblock


\bibitem[Wang et~al\mbox{.}(2020b)]%
        {wang2020massively}
\bibfield{author}{\bibinfo{person}{Xinlei Wang}, \bibinfo{person}{Yuxing Qiu}, \bibinfo{person}{Stuart~R Slattery}, \bibinfo{person}{Yu Fang}, \bibinfo{person}{Minchen Li}, \bibinfo{person}{Song-Chun Zhu}, \bibinfo{person}{Yixin Zhu}, \bibinfo{person}{Min Tang}, \bibinfo{person}{Dinesh Manocha}, {and} \bibinfo{person}{Chenfanfu Jiang}.} \bibinfo{year}{2020}\natexlab{b}.
\newblock \showarticletitle{A massively parallel and scalable multi-GPU material point method}.
\newblock \bibinfo{journal}{\emph{ACM Transactions on Graphics (TOG)}} \bibinfo{volume}{39}, \bibinfo{number}{4} (\bibinfo{year}{2020}), \bibinfo{pages}{30--1}.
\newblock


\bibitem[Wiewel et~al\mbox{.}(2019)]%
        {wiewel2019latent}
\bibfield{author}{\bibinfo{person}{Steffen Wiewel}, \bibinfo{person}{Moritz Becher}, {and} \bibinfo{person}{Nils Thuerey}.} \bibinfo{year}{2019}\natexlab{}.
\newblock \showarticletitle{Latent space physics: Towards learning the temporal evolution of fluid flow}. In \bibinfo{booktitle}{\emph{Computer graphics forum}}, Vol.~\bibinfo{volume}{38}. Wiley Online Library, \bibinfo{pages}{71--82}.
\newblock


\bibitem[Wolper et~al\mbox{.}(2020)]%
        {wolper2020anisompm}
\bibfield{author}{\bibinfo{person}{Joshuah Wolper}, \bibinfo{person}{Yunuo Chen}, \bibinfo{person}{Minchen Li}, \bibinfo{person}{Yu Fang}, \bibinfo{person}{Ziyin Qu}, \bibinfo{person}{Jiecong Lu}, \bibinfo{person}{Meggie Cheng}, {and} \bibinfo{person}{Chenfanfu Jiang}.} \bibinfo{year}{2020}\natexlab{}.
\newblock \showarticletitle{AnisoMPM: Animating Anisotropic Damage Mechanics}.
\newblock \bibinfo{journal}{\emph{ACM Trans. Graph.}} \bibinfo{volume}{39}, \bibinfo{number}{4}, Article \bibinfo{articleno}{37} (\bibinfo{year}{2020}).
\newblock


\bibitem[Wolper et~al\mbox{.}(2019)]%
        {wolper2019cd}
\bibfield{author}{\bibinfo{person}{Joshuah Wolper}, \bibinfo{person}{Yu Fang}, \bibinfo{person}{Minchen Li}, \bibinfo{person}{Jiecong Lu}, \bibinfo{person}{Ming Gao}, {and} \bibinfo{person}{Chenfanfu Jiang}.} \bibinfo{year}{2019}\natexlab{}.
\newblock \showarticletitle{CD-MPM: continuum damage material point methods for dynamic fracture animation}.
\newblock \bibinfo{journal}{\emph{ACM Transactions on Graphics (TOG)}} \bibinfo{volume}{38}, \bibinfo{number}{4} (\bibinfo{year}{2019}), \bibinfo{pages}{1--15}.
\newblock


\bibitem[Xie et~al\mbox{.}(2021)]%
        {xie2021neural}
\bibfield{author}{\bibinfo{person}{Yiheng Xie}, \bibinfo{person}{Towaki Takikawa}, \bibinfo{person}{Shunsuke Saito}, \bibinfo{person}{Or Litany}, \bibinfo{person}{Shiqin Yan}, \bibinfo{person}{Numair Khan}, \bibinfo{person}{Federico Tombari}, \bibinfo{person}{James Tompkin}, \bibinfo{person}{Vincent Sitzmann}, {and} \bibinfo{person}{Srinath Sridhar}.} \bibinfo{year}{2021}\natexlab{}.
\newblock \showarticletitle{Neural Fields in Visual Computing and Beyond}.
\newblock \bibinfo{journal}{\emph{arXiv preprint arXiv:2111.11426}} (\bibinfo{year}{2021}).
\newblock


\bibitem[Xu et~al\mbox{.}(2015)]%
        {xu2015interactive}
\bibfield{author}{\bibinfo{person}{Hongyi Xu}, \bibinfo{person}{Yijing Li}, \bibinfo{person}{Yong Chen}, {and} \bibinfo{person}{Jernej Barbi{\v{c}}}.} \bibinfo{year}{2015}\natexlab{}.
\newblock \showarticletitle{Interactive material design using model reduction}.
\newblock \bibinfo{journal}{\emph{ACM Transactions on Graphics (TOG)}} \bibinfo{volume}{34}, \bibinfo{number}{2} (\bibinfo{year}{2015}), \bibinfo{pages}{1--14}.
\newblock


\bibitem[Yang et~al\mbox{.}(2021)]%
        {yang2021geometry}
\bibfield{author}{\bibinfo{person}{Guandao Yang}, \bibinfo{person}{Serge Belongie}, \bibinfo{person}{Bharath Hariharan}, {and} \bibinfo{person}{Vladlen Koltun}.} \bibinfo{year}{2021}\natexlab{}.
\newblock \showarticletitle{Geometry Processing with Neural Fields}.
\newblock \bibinfo{journal}{\emph{Advances in Neural Information Processing Systems}}  \bibinfo{volume}{34} (\bibinfo{year}{2021}).
\newblock


\bibitem[Yang et~al\mbox{.}(2015)]%
        {yang2015expediting}
\bibfield{author}{\bibinfo{person}{Yin Yang}, \bibinfo{person}{Dingzeyu Li}, \bibinfo{person}{Weiwei Xu}, \bibinfo{person}{Yuan Tian}, {and} \bibinfo{person}{Changxi Zheng}.} \bibinfo{year}{2015}\natexlab{}.
\newblock \showarticletitle{Expediting precomputation for reduced deformable simulation}.
\newblock \bibinfo{journal}{\emph{ACM Trans. Graph}} \bibinfo{volume}{34}, \bibinfo{number}{6} (\bibinfo{year}{2015}).
\newblock


\bibitem[Yue et~al\mbox{.}(2015)]%
        {yue2015continuum}
\bibfield{author}{\bibinfo{person}{Yonghao Yue}, \bibinfo{person}{Breannan Smith}, \bibinfo{person}{Christopher Batty}, \bibinfo{person}{Changxi Zheng}, {and} \bibinfo{person}{Eitan Grinspun}.} \bibinfo{year}{2015}\natexlab{}.
\newblock \showarticletitle{Continuum foam: A material point method for shear-dependent flows}.
\newblock \bibinfo{journal}{\emph{ACM Transactions on Graphics (TOG)}} \bibinfo{volume}{34}, \bibinfo{number}{5} (\bibinfo{year}{2015}), \bibinfo{pages}{1--20}.
\newblock


\bibitem[Yue et~al\mbox{.}(2018)]%
        {yue2018hybrid}
\bibfield{author}{\bibinfo{person}{Yonghao Yue}, \bibinfo{person}{Breannan Smith}, \bibinfo{person}{Peter~Yichen Chen}, \bibinfo{person}{Maytee Chantharayukhonthorn}, \bibinfo{person}{Ken Kamrin}, {and} \bibinfo{person}{Eitan Grinspun}.} \bibinfo{year}{2018}\natexlab{}.
\newblock \showarticletitle{Hybrid grains: Adaptive coupling of discrete and continuum simulations of granular media}.
\newblock \bibinfo{journal}{\emph{ACM Transactions on Graphics (TOG)}} \bibinfo{volume}{37}, \bibinfo{number}{6} (\bibinfo{year}{2018}), \bibinfo{pages}{1--19}.
\newblock


\bibitem[Zehnder et~al\mbox{.}(2021)]%
        {zehnder2021ntopo}
\bibfield{author}{\bibinfo{person}{Jonas Zehnder}, \bibinfo{person}{Yue Li}, \bibinfo{person}{Stelian Coros}, {and} \bibinfo{person}{Bernhard Thomaszewski}.} \bibinfo{year}{2021}\natexlab{}.
\newblock \showarticletitle{Ntopo: Mesh-free topology optimization using implicit neural representations}.
\newblock \bibinfo{journal}{\emph{Advances in Neural Information Processing Systems}}  \bibinfo{volume}{34} (\bibinfo{year}{2021}), \bibinfo{pages}{10368--10381}.
\newblock


\end{thebibliography}

\clearpage
\appendix

\section{Experiment details}
In the bread tearing example, a total of $24$ simulations were generated with varying Young's moduli. A random choice of $4$ simulations were reserved for testing, while the rest $20$ simulations were for training. In the cake-cutting example, the spatula is slicing the cake at different angles. Here $\mathcal{D}_{\text{train}}$ contains $12$ simulations and $\mathcal{D}_{\text{test}}$ contains $3$ simulations. In the example of sand, $20$ simulations for different friction angles, four of which are reserved for testing. In the example of metal, $\mathcal{D}_{\text{train}}$ contains 12 simulations, and $\mathcal{D}_{\text{test}}$ contains 4 simulations. The hardening coefficient $\xi$ is distinct in different simulations. In the example of toothpaste, we also have 12 training simulations and 4 testing simulations. The toothbrush is held at different angles. Finally, the last two examples (jelly cube and squishy ball) both contain 3 training simulations and 1 testing simulation.

\section{Extrapolation, generalization, and training data}
In this section, we provide more aggressive extrapolation and generalization experiments. To demonstrate the differences between testing data (visualized in the main text) and training data, we also visualize the training data in this section.

\begin{figure}[h]
  \centering
  \includegraphics[width=0.9\linewidth]{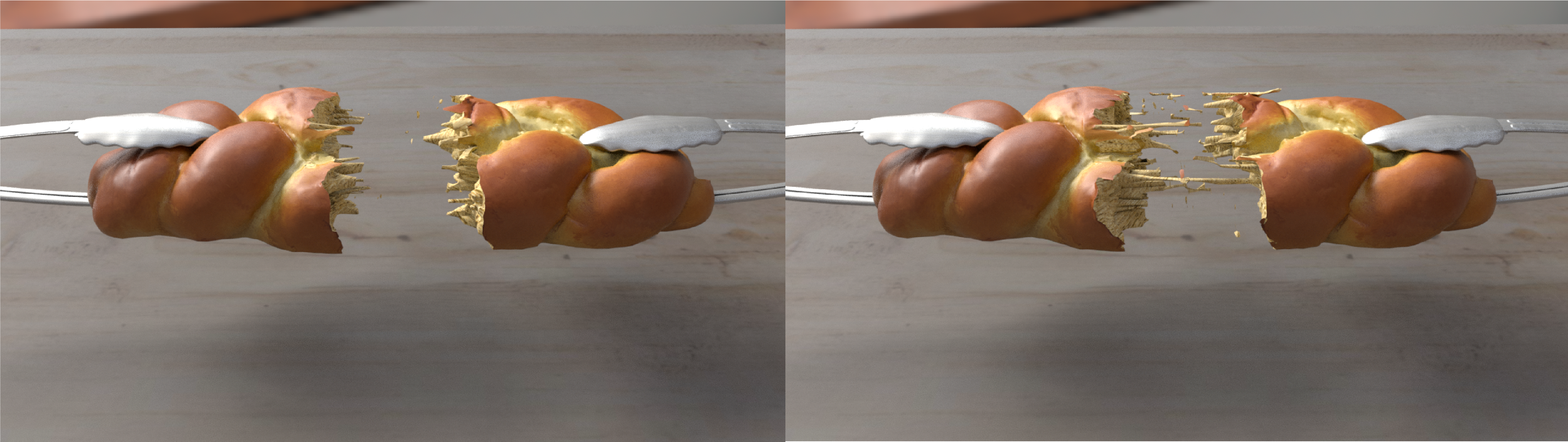}
  \caption{Deployment results with unseen weak elements. \textbf{Left:} A mild perturbation on the weak elements. The simulation remains fairly accurate. \textbf{Right:} A more aggressive perturbation on the weak elements. The simulation suffers from significantly larger errors.}
  \Description{The top right subplots show the corresponding ground truth.}
  \label{fig:unseen_weak}
\end{figure}

In the bread tearing example, the problem parameter is the Young's modulus of the material, and weak elements are inserted to help with fracture. One generalization test is employing unseen weak elements. The results were shown in \cref{fig:unseen_weak}. A mild perturbation on the weak elements (a random perturbation with a scale of $8\%$) would yield a decent result. The total position error is $\delta = 2.7\%.$ On the other hand, if we aggressively perturb the weak elements (a random perturbation with a scale of $30\%$), the resulting deployment will suffer from significant errors. We observe a `partially unsuccessful' fracture. The total position error surges to $\delta = 11\%.$ In summary, our method cannot capture cases where the fracture pattern is drastically different from the ones shown in the training data.

\begin{figure}[h]
  \centering
  \includegraphics[width=0.9\linewidth]{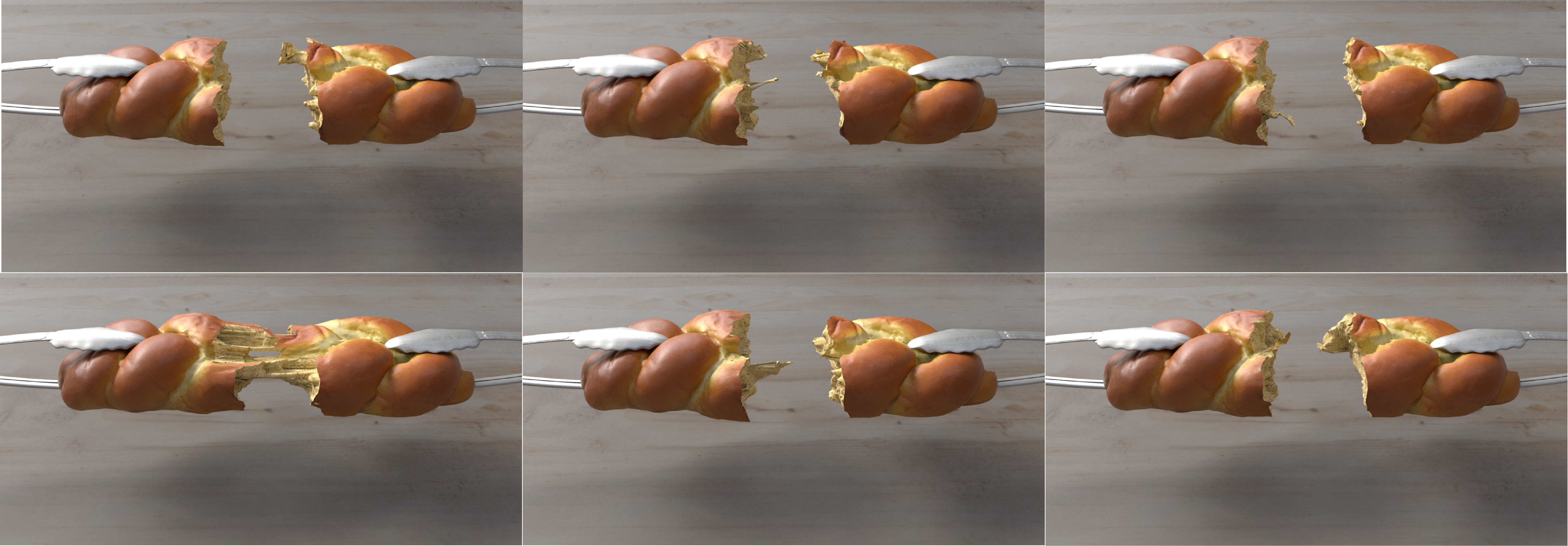}
  \caption{Sample training data of the bread tearing experiment.}
  \Description{The}
  \label{fig:bread_extra_training}
\end{figure}

We also list the training data (\cref{fig:bread_extra_training}) in the bread-tearing experiment, all of which have fracture patterns different from the testing data.

\begin{figure}[h]
  \centering
  \includegraphics[width=0.9\linewidth]{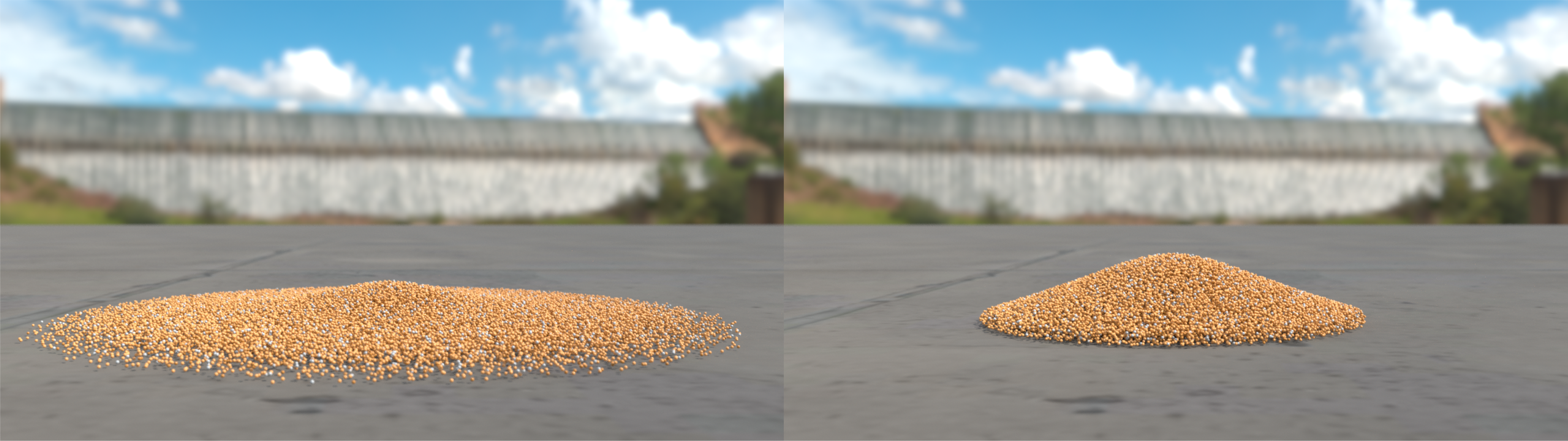}
  \caption{Sample training data for the sand example. \textbf{Left:} simulation with the smallest friction angle  $21^{\circ}$ in the training data set $\mathcal{D}_{\text{train}}.$ \textbf{Right:} simulation with the largest friction angle  $38.5^{\circ}$ in $\mathcal{D}_{\text{train}}.$}
  \Description{The}
  \label{fig:sand_training}
\end{figure}

\begin{figure}[h]
  \centering
  \includegraphics[width=0.9\linewidth]{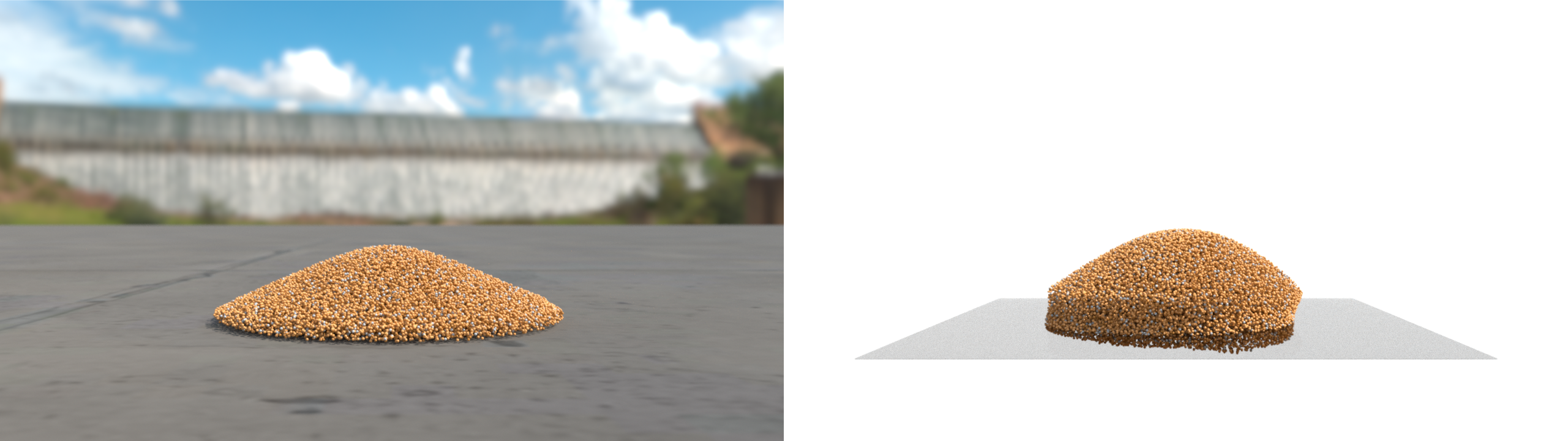}
  \caption{Extrapolation for the sand example. \textbf{Left:} the problem parameter $\theta$ is set to $44^{\circ},$ where $\mathcal{D}_{\text{train}}$ contains several simulations with $\theta \in [21^{\circ}, 38.5^{\circ}].$ The extrapolation yields accurate results, with a total position error of merely $\delta=1.8\%.$ \textbf{Right:} an extreme extrapolation of $\theta = 55^{\circ}$ is performed. Our approach suffers from larger errors.}
  \Description{The}
  \label{fig:sand_extrapolate}
\end{figure}

We also performed several extrapolation tests for the sand experiment. In the training data set $\mathcal{D}_{\text{train}},$ the smallest friction angle is $21^{\circ}$ and the largest is $38.5^{\circ}.$ See c\ref{fig:sand_training}. The testing data shown in the main text are interpolations where the friction angle lies between $21^{\circ}$ and $38.5^{\circ}.$ Our approach is also robust under reasonable extrapolations. For instance, under a friction angle of $44^{\circ},$ our approach generates accurate results, with a total position error of $\delta = 1.8\%$. However, our method will not work under extreme extrapolation. When we set the friction angle to a significantly larger $55^{\circ},$ our latent space dynamics suffer from very larger errors. The sand column neither keeps its symmetry nor obeys its boundary condition (it penetrates the ground). See \ref{fig:sand_extrapolate}. Thus, our current pipeline does not support extreme extrapolation that is significantly different from the training data. 

\section{Runtime analysis}
The timing experiments are performed on a machine with Intel Core i7-8700K and NVIDIA Quadro P6000. Both the full-order and the reduced MPM simulator are implemented in WARP \cite{warp2022}. The neural networks were implemented in PyTorch. The runtime reported is an average of ten repeated trials. 

\begin{figure}[h]
  \centering
  \includegraphics[width=1.0\linewidth]{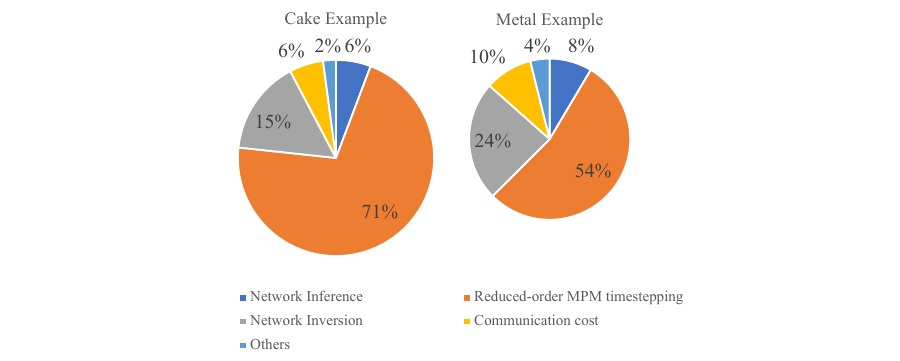}
  \caption{Pie-chart breakdown for the runtime of each component in our reduced-order pipeline for two examples: the cake and the metal. The pie chart for the metal example is plotted smaller to show that this is a smaller problem and thus has a shorter runtime. Overall, the main bottleneck of the reduced-order algorithm remains to be MPM timestepping while the network operations introduce little overhead.}
  \Description{The top right subplots show the corresponding ground truth.}
  \label{fig:breakdown}
\end{figure}

In the example of cake, the runtime of full-order MPM is $4.53s/100$ frames. The runtime of the reduced-order MPM is $0.469s/100$ frames. The cost in network inference, projection, as well as communication between WARP and PyTorch is $0.195/100$ frames. 

In the example of metal, the runtime of full-order MPM is $1.305s/100$ frames. The runtime of the reduced-order MPM is $0.209s/100$ frames. The cost in network inference, projection, as well as communication between WARP and PyTorch is $0.178/100$ frames. 

We also provide two pie charts for the breakdown of the runtime of each component in our reduced-order scheme for the two examples. The runtime for network-related operations grows relatively mildly when problem size increases, whereas the runtime for MPM timestepping in theory grows linearly in problem size (see next paragraph for a discussion). Thus, problems with larger scales tend to enjoy more time savings from our algorithm.

For a paralleled explicit MPM simulator, one can usually observe that the runtime is roughly linearly proportional to the total number of particles. The two majority of costs come from the SVD computation for stress and the atomic addition in P2G. In our experiments, reduced MPM with about $9\%$ of original particles costs about $16\%$ of original runtime, and with $6.7\%$ of original particles costs about $10.3\%$ of original runtime. In theory, this reduced MPM runtime should be even smaller. In addition, SVD is not required in the reduced MPM. We reason that this is because, during deployment, PyTorch has already occupied some memory. Thus, the computing power allocated for the reduced MPM solver in WARP would be less than that allocated for a full-order MPM solver alone.

\section{Training Details}
The training data is generated from an MPM solver written in WARP \cite{warp2022} under double precision. The Adam optimizer \cite{kingma2014adam} for stochastic gradient descent is used for training. The Xavier initialization is used for the ELU layers. We fix the learning rates to be $(10^{-3}, 5 \times 10^{-4}, 2 \times 10^{-4}, 10^{-4}, 5 \times 10^{-5}).$ For the Neural Deformation Field, $300$ epochs are trained for the learning rate above. For the Neural Stress Field and Neural Affine Field, $600$ epochs are trained for the learning rate above. The batch size for the three manifolds is $k \cdot |\mathcal{P}|$ where we choose $k$ to be the largest value such that the training data fits within memory or $32,$ whichever is smaller. The training data is normalized to have mean zero and variance one. The whole training pipeline is implemented in Pytorch Lightning \cite{pytorchlightning}.
\section{Network architecture}
The (input, output) dimension pairs for $\boldsymbol{g}, \boldsymbol{h}$ and $\boldsymbol{l}$ are $(3+r, 3), (3+r, 6), $ and $(3+r, 9),$ respectively. The Kirchhoff stress $\boldsymbol{\tau}$ has $6$ degrees of freedom since it is symmetric.
Each MLP network contains $5$ hidden layers, each of which has a width of $\beta \cdot d$, where $d=3$ for $\boldsymbol{g},$ $d=6$ for $\boldsymbol{h},$ and $d=9$ for $\boldsymbol{l}.$ $\beta$ is a hyperparameter, the exact value of which for each experiment is listed in Table 1 in the main text. We adopt the ELU activation function \cite{clevert2015fast}.
The encoder network is devised as the following: several 1D convolution layers of kernel size 6, stride size 4, and output channel size 3 are applied until the length of the 1D output vector reaches or below 12. The vector is then reshaped to 1 channel. One MLP layer transforms its dimension to 32, followed by the last MLP layer that outputs a vector with dimension $r.$
\section{Elasticity and plasticity details}
We first list all parameters that shall be needed in discussing the models below.
$$\begin{array}{lll}
\hline \text { Notation } & \text { Meaning } & \text { Relation to }(E, \nu) \\
\hline E & \text { Young's modulus } & / \\
\nu & \text { Poisson's ratio } & / \\
\hat{\mu} & \text { Shear modulus } & \hat{\mu}=\frac{E}{2(1+\nu)} \\
\lambda & \text { Lamé modulus } & \lambda=\frac{E \nu}{(1+\nu)(1-2 \nu)} \\
\kappa & \text{Bulk modulus} & \kappa = \frac{E}{3(1-2 \nu)}\\
\hline
\end{array}$$
In all plasticity models used in our work, the deformation gradient is multiplicatively decomposed into $\boldsymbol{F}=\boldsymbol{F}^E \boldsymbol{F}^P$ following some yield stress condition. A hyperelastic constitutive model is applied to $\boldsymbol{F}^E$ to compute the Kirchhoff stress $\boldsymbol{\tau}.$ For a pure elastic continuum, one simply takes $\boldsymbol{F}^E=\boldsymbol{F}.$

\subsection{Fixed corotated elasticity}
The Kirchhoff stress $\boldsymbol{\tau}$ is defined as 
\begin{equation}
\boldsymbol{\tau} = 2 \hat{\mu}(\boldsymbol{F}^E-\boldsymbol{R}) {\boldsymbol{F}^E}^{T}+\lambda(J-1) J,
\end{equation}
where $\boldsymbol{R} = \boldsymbol{U} \boldsymbol{V}^T$ and $\boldsymbol{F}^E = \boldsymbol{U}\boldsymbol{\Sigma}\boldsymbol{V}^T$ is the singular value decomposition of elastic deformation gradient. \cite{jiang2015affine}
\subsection{StVK elasticity}
The Kirchhoff stress $\boldsymbol{\tau}$ is defined as 
\begin{equation}
    \boldsymbol{\tau}=\boldsymbol{U} \left(2\hat{\mu} \boldsymbol{\epsilon}+\lambda \operatorname{sum}(\boldsymbol{\epsilon}) \mathbf{1}\right) \boldsymbol{V}^T,
\end{equation}
where $\boldsymbol{\epsilon}=\log (\boldsymbol{\Sigma})$ and $\boldsymbol{F}^E = \boldsymbol{U}\boldsymbol{\Sigma}\boldsymbol{V}^T.$ \cite{klar2016drucker}
\subsection{Drucker-Prager plasticity}
The return mapping of Drucker-Prager plasticity for sand \cite{klar2016drucker} is, given $\boldsymbol{F} = \boldsymbol{U}\boldsymbol{\Sigma}\boldsymbol{V}^T$ and $\boldsymbol{\epsilon}=\log (\boldsymbol{\Sigma}),$ $$\boldsymbol{F}^E = \boldsymbol{U} \mathcal{Z}(\boldsymbol{\Sigma}) \boldsymbol{V}^T.$$
$$
\mathcal{Z}(\boldsymbol{\Sigma})=\left\{\begin{array}{ll}
\mathbf{1} & \operatorname{sum}(\boldsymbol{\epsilon})>0 \\
\boldsymbol{\Sigma} & \delta \gamma \leq 0, \text { and } \operatorname{sum}(\boldsymbol{\epsilon}) \leq 0 \\
\exp \left(\boldsymbol{\epsilon}-\delta \gamma \frac{\hat{\epsilon}}{\|\hat{\epsilon}\|}\right) & \text { otherwise }
\end{array}\right.
$$
Here $\delta \gamma=\|\hat{\boldsymbol{\epsilon}}\|+\alpha \frac{(d \lambda+2 \hat{\mu}) \operatorname{sum}(\boldsymbol{\epsilon})}{2 \hat{\mu}},$ $\alpha=\sqrt{\frac{2}{3}} \frac{2 \sin \phi_f}{3-\sin \phi_f},$ and $\phi_f$ is the friction angle. $\hat{\epsilon} = \operatorname{dev}(\epsilon)$
\subsection{von Mises plasticity}
Given $\boldsymbol{F} = \boldsymbol{U}\boldsymbol{\Sigma}\boldsymbol{V}^T$ and $\boldsymbol{\epsilon}=\log (\boldsymbol{\Sigma}),$ $$\boldsymbol{F}^E = \boldsymbol{U} \mathcal{Z}(\boldsymbol{\Sigma}) \boldsymbol{V}^T,$$
where $$
\mathcal{Z}(\boldsymbol{\Sigma})=\left\{\begin{array}{ll}
\boldsymbol{\Sigma}, & \left\|\boldsymbol{\tau}-\frac{1}{d} \operatorname{sum}(\boldsymbol{\tau})\right\|-\tau_y \leq 0 \\
\exp \left(\boldsymbol{\epsilon}-\delta \gamma \frac{\hat{\epsilon}}{\|\boldsymbol{\epsilon}\|}\right), & \text { Otherwise }
\end{array}\right.,
$$
and $\delta \gamma=\|\hat{\boldsymbol{\epsilon}}\|_F-\frac{\tau_y}{2 \hat{\mu}}.$ Here $\tau_Y$ is the yield stress. If hardening is included, the yield stress is updated as $\tau_Y^{n+1}=\tau_Y^n+2 \hat{\mu} \xi \delta \gamma,$ where $\xi$ is the hardening coefficient. If softening is included, yield stress is updated as $\tau_Y^{n+1}=\tau_Y^n -\theta\|\boldsymbol{\epsilon}-\operatorname{proj}(\boldsymbol{\epsilon})\|_F.$ When $\tau_Y$ reaches zero, the material is considered damage and its Lamé parameters are set to zero. \cite{wang2019simulation} 
\subsection{Herschel-Bulkley plasticity}
We follow \cite{yue2015continuum} and take the simple case where $h=1.$ Denote $\boldsymbol{s}^{\text{trial}} = \operatorname{dev}(\boldsymbol{\tau}^{\text{trial}}),$ and $s^{\text{trial}} = ||\boldsymbol{s}^{\text{trial}}||.$ The yield condition is $\Phi(s)=s-\sqrt{\frac{2}{3}} \sigma_Y \leq 0.$ If it is violated, we modify $s^{\text{trial}}$ by $$
s=s^{\text{trial}}-\left(s^{\text{trial}}-\sqrt{\frac{2}{3}} \sigma_Y\right) /\left(1+\frac{\eta}{2 \hat{\mu} \Delta t}\right).
$$
$\boldsymbol{s}$ can then be recovered as $\boldsymbol{s} = s \cdot \frac{\boldsymbol{s}^{\text{trial}}}{||\boldsymbol{s}^{\text{trial}}||}.$
Define $\boldsymbol{b}^E = \boldsymbol{F}^E {\boldsymbol{F}^E}^T.$ The Kirchhoff stress $\boldsymbol{\tau}$ is computed as 
$$\boldsymbol{\tau} =\frac{\kappa}{2}\left(J^2-1\right) \boldsymbol{I}+\hat{\mu} \operatorname{dev}\left[\operatorname{det}(\boldsymbol{b}^E)^{-\frac{1}{3}} \boldsymbol{b}^E\right].$$

\clearpage
\end{document}